\documentclass[12pt]{IOPART}
\usepackage{graphicx}
\def\lbets{$\lambda$-(BETS)$_{2}$GaCl$_{4}$}
\def\hc2{$H_{c2}$}

\def\cuscn{$\kappa$-(BEDT-TTF)$_2$Cu(NCS)$_2$}

\def\nh4{$\alpha$-(BEDT-TTF)$_2$NH$_4$Hg(SCN)$_4$}

\begin{document}

\title[Fermiology and superconductivity of \lbets]{Superconducting
properties and Fermi-surface topology of the quasi-two-dimensional
organic superconductor
$\lambda$-(BETS)$_{2}$GaCl$_{4}$.}
\author{Charles Mielke$^1$,\footnote[3]{To
whom correspondence should be addressed (cmielke@lanl.gov)}
John Singleton$^{1,2}$, Moon-Sun Nam$^2$,
Neil Harrison$^1$, C.C. Agosta$^3$, B. Fravel$^4$ and L. K. Montgomery$^4$}

\address{$^1$National High Magnetic Field Laboratory,
Los Alamos National Laboratory, MS-E536,
Los Alamos, New Mexico 87545, USA}

\address{$^2$University of Oxford, Department of Physics, The
Clarendon Laboratory, Parks Road,
Oxford OX1 3PU, U.K.}

\address{$^3$Department of Physics, Clark University, Worcester,
MA 01610, USA}

\address{$^4$Department of Chemistry,
Indiana University, Bloomington, Indiana 47405, USA}

\begin{abstract}
The Fermi surface topology of the organic
superconductor \lbets ~has been
determined using the Shubnikov-de Haas and magnetic breakdown
effects and angle-dependent magnetoresistance oscillations.
The former experiments were carried out in pulsed fields
of up to 60~T, whereas the latter employed quasistatic fields of up to 30~T.
All of these data show that the Fermi-surface topology of
\lbets is very similar to that of the most heavily-studied
organic superconductor, \cuscn , except in one important
respect; the interplane transfer integral in \lbets
~is a factor $\sim 5$ larger than that in \cuscn .
The increased three-dimensionality of \lbets
~is manifested
in radiofrequency penetration-depth measurements,
which show a clear dimensional crossover in the behaviour
of $H_{\rm c2}(T)$. The radiofrequency measurements have also
been used to extract the Labusch parameter determining
the fluxoid interactions as a function of temperature,
and to map the flux-lattice melting curve.
\end{abstract}


\submitto{\JPCM}

\maketitle

\section{Introduction}
There is considerable current debate over the
nature of superconductivity in quasi-two-dimensional (Q2D) crystalline
organic metals~\cite{review,carrington,elsinger,tunnelling}.
The most heavily studied members of this family of materials
are the $\kappa$-phase BEDT-TTF salts (e.g. \cuscn)~\cite{review}.
Whilst nuclear magnetic resonance~\cite{french}, penetration-depth~\cite{carrington},
tunnelling~\cite{tunnelling} and other experiments~\cite{review} appear
to suggest that the superconductivity in these salts
may be d-wave-like and mediated by
spin-density-wave-like fluctuations, some doubts have been cast
by recent controversial specific heat measurements, which may
suggest that the order parameter
does not possess the required nodes~\cite{elsinger}.
Several theories~\cite{schmalian,aoki,maki,charffi} stress
the importance of the details of the Fermi-surface topology
in providing suitable prerequisites for superconductivity;
if the Fermi-surface geometry and interactions
are altered slightly, it appears that BCS-like s-wave superconductivity {\it may}
be the dominant low-temperature groundstate~\cite{review}.

Clearly, it is of importance to study organic superconductors
with slight variations in Fermi-surface topology
so that the
effect on the superconducting groundstate can be assessed.
In this paper we therefore report magnetotransport
and radiofrequency penetration-depth measurements of the
superconductor \lbets . Shubnikov-de Haas oscillations,
magnetic breakdown and angle-dependent magnetoresistance
oscillations indicate that
the effective masses of
\lbets ~and much of the topology of its Fermi surface
are similar to those of
the most heavily-studied BEDT-TTF superconductor, \cuscn.
However, the magnetoresistance close to $\theta =90^{\circ}$
implies that the interplane transfer integral in
\lbets ~is approximately five times larger than that
in \cuscn , suggesting that
\lbets ~is less two-dimensional.
This increased dimensionality
is manifested in the superconducting properties of
\lbets ; a clear two-dimensional to three dimensional
crossover is seen in the temperature dependence of $H_{\rm c2}$.
We have also used the radiofrequency measurements to extract
various parameters related to the interactions between the fluxoids
and to reveal the melting of the vortex solid.

\section{Background information}
When the innermost four sulphur atoms of BEDT-TTF~\cite{review}
are replaced by
selenium to produce BETS (where BETS stands for
bis(ethylenedithio)tetraselenafulvalene), electrocrystallization
gives rise to a range of charge-transfer salts with notably
different properties compared to their BEDT-TTF
counterparts~\cite{betsreview1,growthdetails,calculation,grow2}. Salts
of the $\lambda$-phase morphology are presently unique to the
BETS series, and \lbets ~($T_{\rm c}\approx 5$~K)
remains the only superconducting BETS charge-transfer salt
found thus far~\cite{calculation,ujiuji}.

Crystals of the $\lambda$-phase exist in the form of needles with
the long axis of the needle corresponding to the shortest lattice
vector ${\bf c}$~\cite{growthdetails,calculation}.
At first sight, the crystal structure looks
quasi-one-dimensional, with the BETS molecules packing
roughly parallel in the planes between the anions~\cite{calculation}.
However, the BETS sites are not equivalent, and
the cation molecules in fact
occur in dimers, surrounded roughly isotropically
by four nearest-neighbour dimers with the same orientation.
The crystallographic unit cell contains two dimers,
and hence contributes two holes~\cite{calculation}.
Although the details of the cation positioning and symmetry
are rather different than in \cuscn ~\cite{review}, the
overall similarity of the dimer arrangements lead
one to expect a Fermi surface for \lbets
~which is topologically similar to that in \cuscn .
Indeed, the calculated bandstructure of \lbets ~predicts a
Fermi surface consisting of a
quasi-two-dimensional (Q2D) hole pocket (the $\alpha$ pocket) and
a pair of warped quasi-one-dimensional
(Q1D) sheets~\cite{calculation,whangbo}.
According to the calculations, the $\alpha$ pocket in
this case is expected to occupy $\sim 28-33$~\% of the Brillouin
zone~\cite{calculation,whangbo}.

\section{Magnetotransport studies.}
\label{bands}
\subsection{Experimental details.}
Single crystals of approximately $1\times 0.1\times 0.05$~mm$^3$ were
synthesized using electrochemical techniques~\cite{calculation} employing
a 1,1,-trichloroethane/1,1,2-trichloroethane/ethanol solvent system~\cite{montya}.
For the purpose
of performing four-wire resistance measurements, $12~\mu$m gold
leads were attached to the samples using
graphite paint. In the pulsed field experiments,
the resistance was measured using a $10~\mu$A ac current
with a frequency of 200~kHz~\cite{review,pulsed}. The voltage
was measured using a high-speed lock-in amplifier.
Temperatures as low as $\sim$ 340 mK
were achieved by immersing the sample in
liquid $^3$He inside a plastic cryostat~\cite{pulsed}.
Capacitor-driven, $\sim 40$~millisecond-duration pulsed magnetic fields
of up to 60~T were provided by the National High
Magnetic Field Laboratory (NHMFL), Los Alamos.
Angle-dependent magnetoresistance (AMRO) studies were
made using a two-axis rotation insert~\cite{review}
in quasistatic magnetic fields
of up to 30~T provided by NHMFL, Tallahassee.
In the AMRO experiments, an ac current of $5~\mu$A
(frequency $30-80$~Hz) was used for the resistance measurements, and
a stable base temperature of 1.4~K was obtained
by pumping on $^4$He liquid.
In both pulsed and quasistatic measurements, the current through
the sample was driven in the interplane ${\bf b}^*$ direction;
in such a configuration the measured resistance is accurately
proportional to the interplane
resistivity component $\rho_{zz}$~\cite{review}.
\subsection{Pulsed-field magnetotransport: Shubnikov-de Haas and magnetic breakdown oscillations.}
\begin{figure}[htbp]
\centering
\includegraphics[height=10cm]{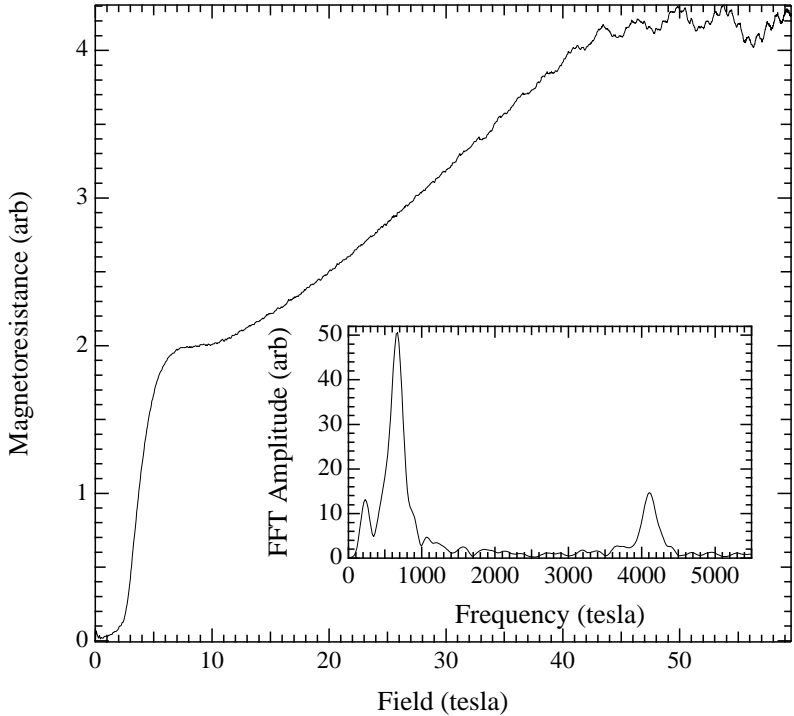}
\caption{Resistance of an \lbets ~crystal
as a function of magnetic field,
applied parallel to the ${\bf b}^*$ direction;
the temperature is 340~mK.
The inset shows a Fourier transform of the data
after subtraction of the non-oscillatory beackground~\cite{review}.
The peak at 650~T is associated with
the $\alpha$ pocket; that at 4030~T is
due to the breakdown ($\beta$) orbit which encompasses 100 \% of the
Brillouin zone.}
\label{mr}
\end{figure}

Figure~\ref{mr} shows the magnetoresistance
of a \lbets ~crystal obtained using pulsed magnetic fields;
the temperature was 340~mK and the magnetic field
was applied parallel to the ${\bf b}^*$ direction
({\it i.e.} perpendicular to the Q2D planes of the crystal)~\cite{growthdetails}.
After the superconducting to normal transition,
the resistance risies until a
series of low-frequency Shubnikov-de Haas oscillations
emerges at about 33~T.
These grow in amplitude, until at about 45~T, a higher frequency
series of oscillations becomes visible.
The inset shows a Fourier transform of the magnetoresistance data.
The lower of the two frequencies $F_{\alpha}$ (believed
to originate from the $\alpha$ pocket) is $650\pm 5$~T. The higher
frequency of $4030~\pm 25$~T, which occurs at fields above
$\sim$~45~T, corresponds to an area in $k$-space approximately
equal to the Brillouin-zone cross-section;
following common usage in other charge-transfer
salts~\cite{review},
we will refer to this as the $\beta$ frequency $F_\beta$.
Frequencies equivalent to the Brillouin-zone area
are readily observed in other
charge-transfer salts (typically in $\kappa$ and
$\alpha$-phase salts of the form
(BEDT-TTF)$_2$X) as a result of magnetic breakdown,
whereby electrons tunnel between the Q1D and Q2D sections of the
Fermi surface~\cite{review}.

The observation of magnetic breakdown is not
unexpected in this material, given the small size of the gap
between the Q2D and Q1D Fermi-surface sections
predicted by the bandstructure calculations~\cite{calculation}.
However,
the experimentally-observed value of $F_{\alpha} \approx 650$~T
is roughly a factor two smaller than that predicted by the
calculations.
Discrepancies between model and experiment
of this size are not unknown in
charge-transfer salts~\cite{house,msamro}.
\subsection{Effective mass determination.}
Although the Fermi surface of \lbets ~is Q2D,
the relatively small amplitudes of the
oscillations in the magnetoresistance (see Figure~\ref{mr}),
and the absence of harmonics in the Fourier transform (inset
to Figure~\ref{mr}) imply that
the Lifshitz-Kosevich (LK) theory
should provide an accurate description of the temperature
dependence of the oscillations~\cite{review}.
According to this theory, the
thermal damping factor has the form~\cite{shoenberg}
\begin{equation}
   R_{T}(B,T)=
   \frac{\chi(\frac{m^{\ast}}{m_{\rm e}})\frac{T}{B}}
   {\sinh{(\chi(\frac{m^{\ast}}{m_{\rm e}})\frac{T}{B})}},
   \label{LKeq}
\end{equation}
where $\chi = 14.69$~TK$^{-1}$ and $m_{\rm e}$ is the free
electron mass. On fitting the amplitudes of the oscillations
~as a function of temperature
(14 different temperatures ranging from 340~mK to 3.0~K were used)
we obtain the effective masses
$m_\alpha^\ast =3.6 \pm 0.1~m_{\rm e}$ for $F_\alpha$ and
$m_\beta^\ast\ =6.3 \pm 1~m_{\rm e}$ for $F_\beta$.

Table~\ref{kappatab} compares the effective masses and Fermi-surface areas
obtained in \lbets ~and the most heavily-studied
$\kappa$-phase BEDT-TTF superconductor, \cuscn ~\cite{caulfield}
~(see Section~3.2 of Reference~\cite{review} for similar data
on other $\kappa$-phase BEDT-TTF salts).
Note that the Fermi-surface parameters of the two salts are remarkably
similar. Moreover, bandstructure calculations in both salts
predict $m^*_{\alpha} \sim m_{\rm e}$~\cite{calculation,caulfield},
whereas the observed
masses are a factor $\sim 3.5$ bigger than this, indicating that
interactions which renormalise the quasiparticle
masses~\cite{review,caulfield,QBB}
are of similar importance in both materials.
In both cases, the bandstructure calculations~\cite{calculation,caulfield}
predict that
$m^*_{\beta} \approx 2m^*_{\alpha}$,
in reasonable agreement with the experimental
values, and suggesting that the renormalising
interactions influence both Q1D and Q2D Fermi-surface
sections in a similar manner~\cite{review,caulfield}.
\begin{table}[tbp] \centering
\begin{tabular}{|l|l|l|l|l|l|l|l|}
\hline
 Salt &  $m^*_{\alpha}/m_e$ & $F_{\alpha}$~(T) & $m^*_{\beta}/m_e$ &
$F_{\beta}$~(T) & $T_{\rm c}$ (K) & Source \\
\hline
\cuscn & 3.5 & 600 & 6.5 & 3920 & 10.4 & \cite{caulfield} \\ \hline
\lbets & 3.6 & 650 & 6.3 & 4030 & 5 & present work \\ \hline
\end{tabular}
\caption{Comparison of magnetic quantum
oscillation frequencies and effective masses
in \lbets ~ and \cuscn .
The table shows the effective masses $m^*_{\alpha}$
and $m^*_{\beta}$ and
frequencies $F_{\alpha}$ and $F_{\beta}$ corresponding to
the $\alpha$ and $\beta$ orbits of the Fermi surface.}
\label{kappatab}
\end{table}
The only marked difference between \lbets
~and \cuscn ~is the relatively high value
of the Dingle Temperature in the former material;
the rate of growth of the oscillations with increasing field
(see Figure~1) suggests a Dingle temperature of
$T_{\rm D}\ \approx 3.2 \pm 0.1$~K for the $\alpha$ pocket.
Typical values of $T_{\rm D}$ in \cuscn ~crystals~\cite{review,tossnitza}
~(and indeed in $\kappa$-phase BETS salts~\cite{kappabets}) are often
a factor $\sim 5$ smaller than this, indicating that
the impurity scattering rate in \lbets ~is relatively high.
The reason for this difference is not yet clear.

\subsection{Angle-dependent magnetoresistance oscillations (AMROs).}
\begin{figure}[htbp]
\centering
\includegraphics[height=10cm]{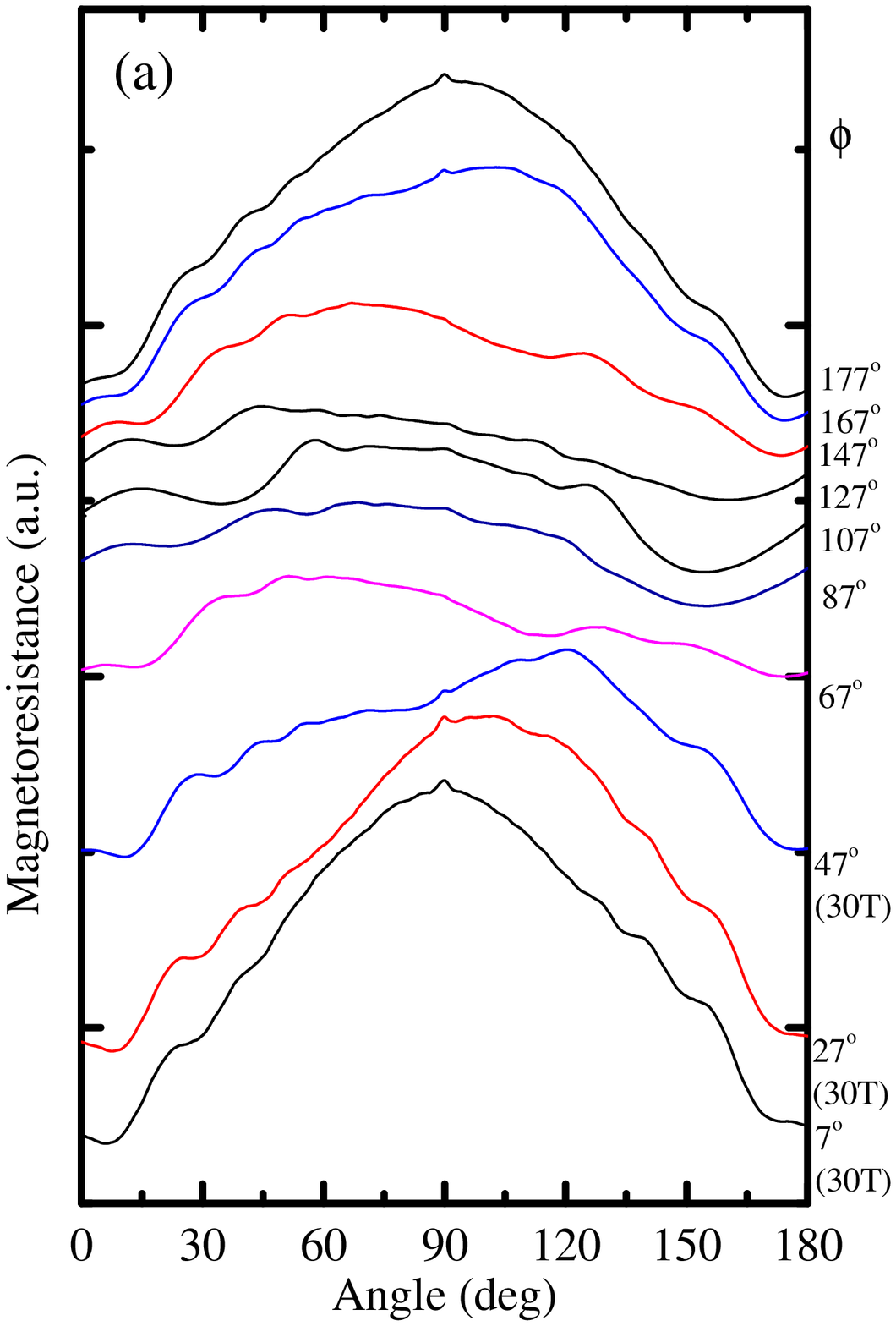}
\includegraphics[height=10cm]{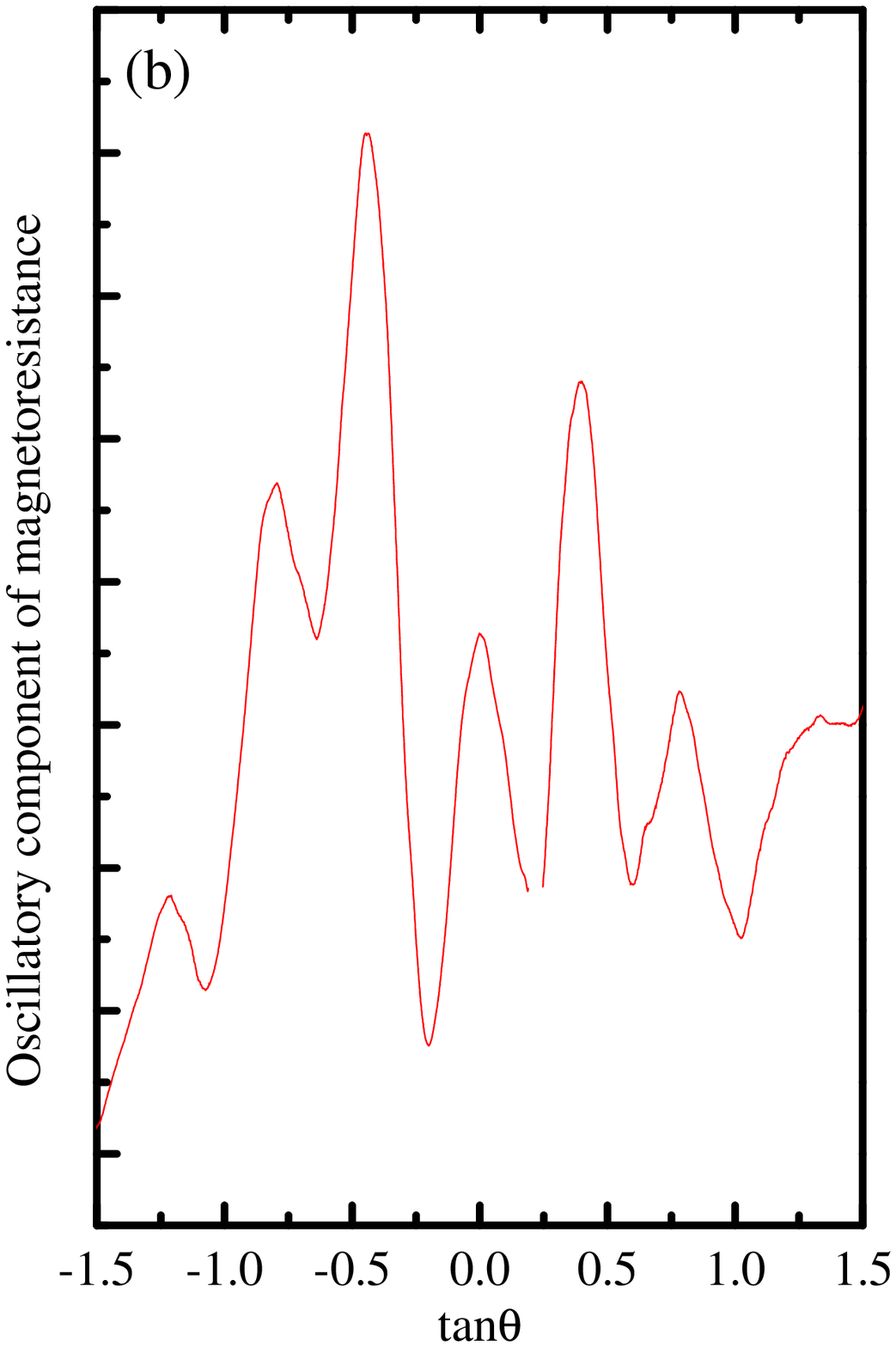}
\caption{(a)~Magnetoresistance of an \lbets ~crystal as a function of $\theta$,
the angle between the applied magnetic field
and ${\bf b}^*$. Data are shown for a number of
different $\phi$ angles (listed at right side of the
Figure), where $\phi$ is azimuthal angle between
the plane of rotation and the ${\bf b}^*{\bf c}$ plane.
The lowest three traces were recorded at 30~T;
the rest of the data were acquired at 27~T.
AMROs are observed as gentle oscillations of the
resistance, periodic in $\tan \theta$;
the small peak at $90^{\circ}$ is due to the presence of
a small number of
closed quasiparticle orbits on the warped
Fermi-surface sections.
(b)~Illustration of the method of locating the AMRO resistance
maxima ($\phi=7^{\circ}$ data from (a)). The slowly-varying background
magnetoresistance has been fitted to a fourth-order polynomial in $\theta$
and subtracted from the experimental data, leaving the oscillatory
component. Peaks periodic in $\tan \theta$ are plainly visible.}
\label{amro}
\end{figure}

\begin{figure}[htbp]
\centering
\includegraphics[height=6cm]{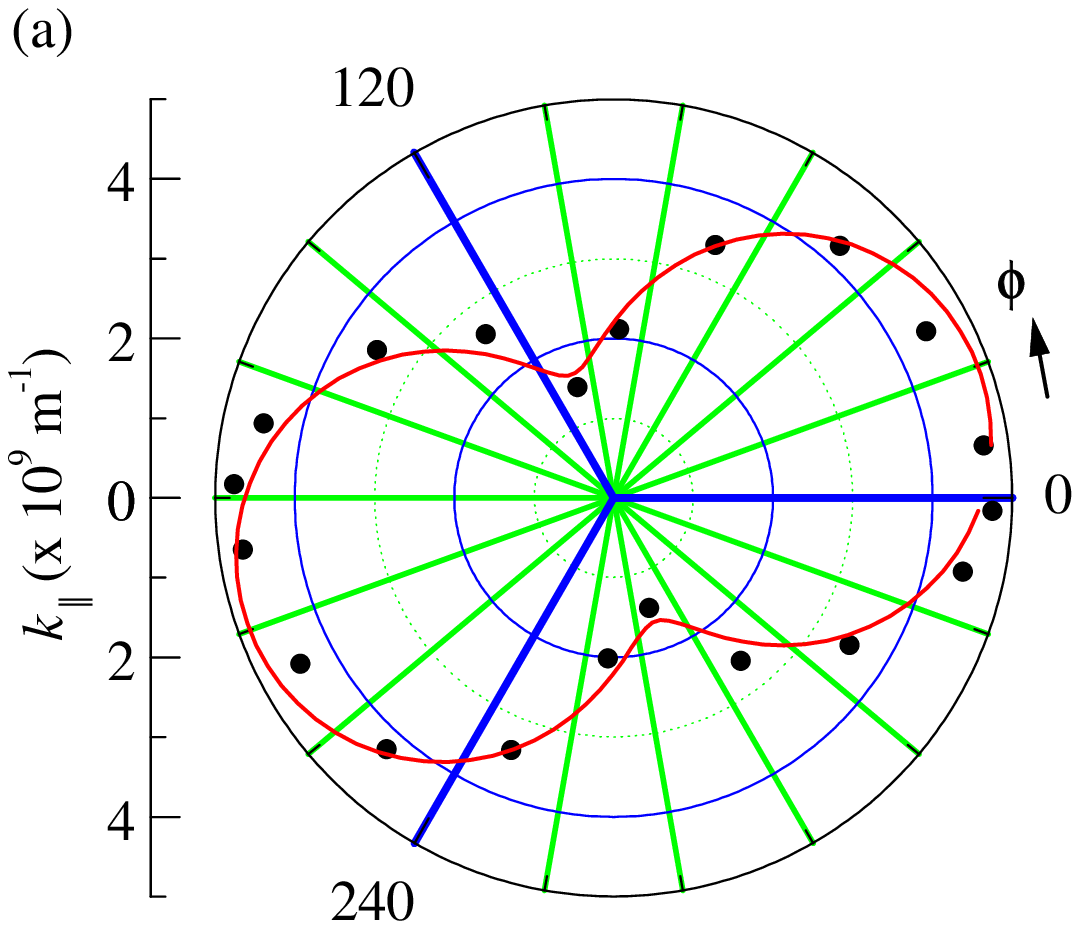}
\includegraphics[height=6cm]{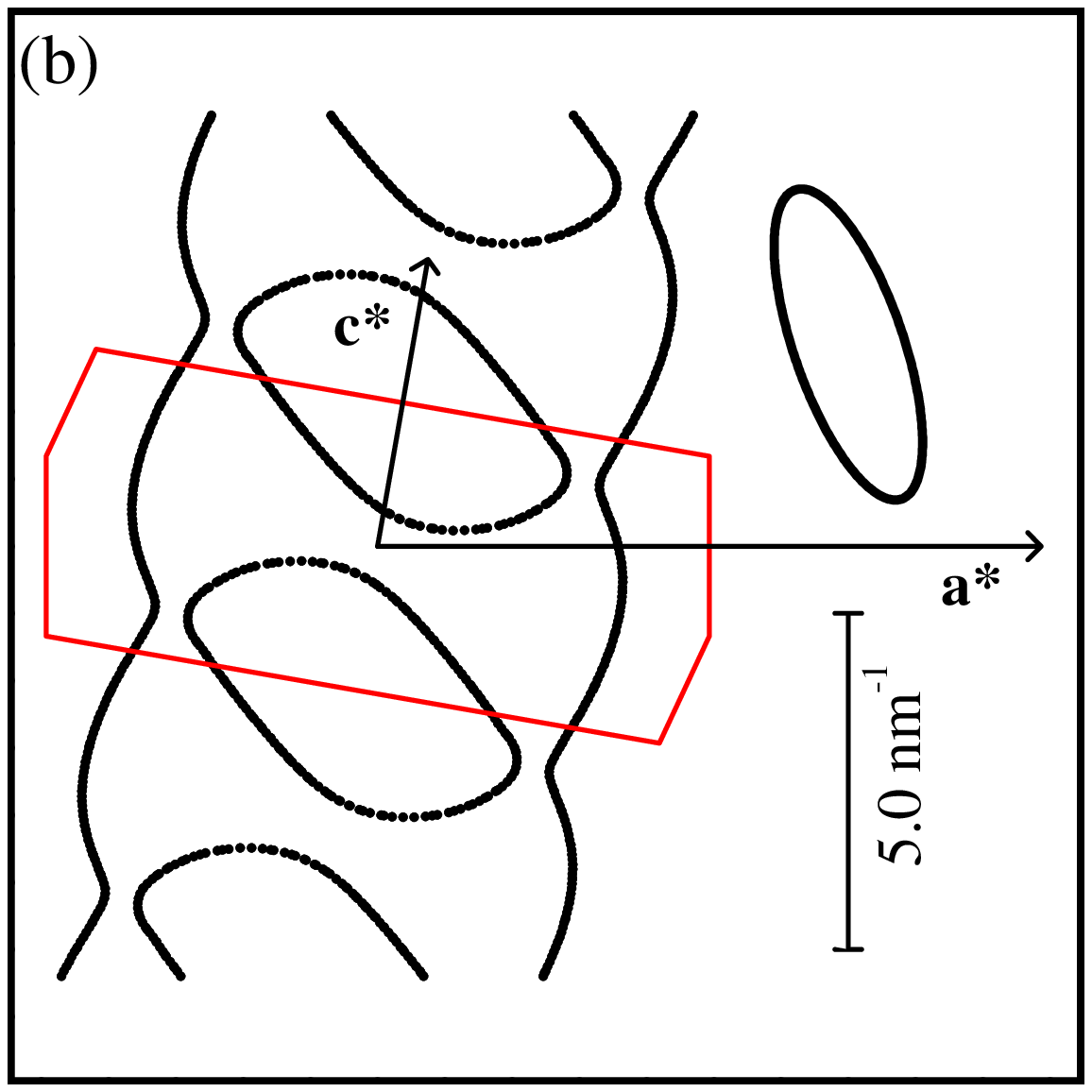}
\caption{(a)~Locus of $k_\|$ versus azimuthal angle
$\phi$ derived from fits of Equation~\ref{amro1}
to the AMRO data. Data are points
and the curve is a fit to Equation~\ref{locus}
with the parameters
$k_x=4.86 \pm 0.08~$nm$^{-1}$,
$k_y=1.63 \pm 0.01~$nm$^{-1}$, $\xi_{\rm inc}= 19 \pm 5^\circ$
and $b^\prime =18.4$~\AA.
(b)~The experimental cross-sectional shape
and orientation of the Q2D Fermi-surface pocket,
shown alongside
(and to the same scale as) the Brillouin zone,
reciprocal lattice vectors ${\bf a}^*$, ${\bf c}^*$
and calculated Fermi surface of Reference~\cite{whangbo}.
The pocket is described by the parameters
$k_x=2.43 \pm 0.04~$nm$^{-1}$,
$k_y=0.815\pm 0.005~$nm$^{-1}$, $\xi_{\rm inc}= 19 \pm 5^\circ$
and $b^\prime =2\times 18.4$~\AA~$\approx 36.8$~\AA
(see text and Equation~\ref{locus}).
}
\label{locust}
\end{figure}

Figure \ref{amro}(a) shows the magnetoresistance
of a \lbets ~crystal as a function of $\theta$,
the angle between the applied magnetic field
and ${\bf b}^*$. Data are shown for a number of
different $\phi$ angles, where $\phi$ is azimuthal angle between
the plane of rotation and the ${\bf b}^*{\bf c}$ plane.
Distinct AMROs are observed; their $\phi$ dependence
suggests that they are caused by a Q2D Fermi-surface
section~\cite{review,msamro}.
In such a case, maxima in the magnetoresistance
occur at angles $\theta_i$ defined by~\cite{review,msamro}
\begin{equation}
   b^\prime k_\| \tan{\theta_i}=\pi(i\pm\frac{1}{4})+A(\phi),
   \label{amro1}
\end{equation}
where $i$ is an integer, $k_\|$ is
the maximum Fermi wave-vector projection on the plane of rotation
of the field and $b^\prime$ is the effective interplane spacing
(see Fig.~\ref{amro}(b)). On
plotting the positions of the maxima $\theta_i$ versus $i$, taking
account of the correct sign of the $\frac{\pi}{4}$
term~\cite{house,msamro}, we obtain
straight lines at all azimuthal angles in accordance with these
expectations. On choosing $b^\prime$ to be
interlayer spacing ($18.4~$\AA) obtained from X-ray
diffraction studies~\cite{calculation,whangbo},
we obtain the locus for $k_\|$ versus $\phi$
shown in Figure~\ref{locust}(a).

A locus in the shape of a figure of
eight is the usual result for a pocket
of elliptical cross-section~\cite{review,house,msamro}.
For a pocket
of ideal elliptical geometry, the locus of $k_\|$ is given by
\begin{equation}
   k_\|=[k_x^2\cos^2(\phi-\xi_{\rm inc})+k_y^2\sin^2(\phi-\xi_{\rm inc})]^\frac{1}{2},
   \label{locus}
\end{equation}
where $\xi_{\rm inc}$ is the inclination of the major axis of the ellipse
with respect to ${\bf b}^*{\bf c}$ plane. The parameters
$k_x=4.86 \pm 0.08~$nm$^{-1}$,
$k_y=1.63 \pm 0.01~$nm$^{-1}$ and $\xi_{\rm inc}= 19 \pm 5^\circ$
yield the best fit (solid curves in Figure~\ref{locust}(a)).

The values of $k_x$ and $k_y$ deduced from
the AMRO data indicate an ellipse  area
corresponding to a Shubnikov-de Haas frequency of
$2608 \pm 60$~T. This is almost exactly four times larger than the
value of $F_{\alpha}$ observed in \lbets
~(see Table~\ref{kappatab}).
It is
inconceivable that this value of $\sim$~2608~T could be the actual
area of the $\alpha$ pocket, as it would then occupy $\sim$~66~\%
of the Brillouin zone.
Quantum oscillation measurements remain the
definitive method for obtaining Fermi-surface cross-section
areas~\cite{review,shoenberg}.
On the other hand, the AMRO measurements shown in
Figure~\ref{amro} behave exactly as one would expect for a Q2D
Fermi-surface pocket~\cite{msamro},
with no evidence for any significant
misalignment of the sample; {\it i.e.} the fits to Equation~\ref{amro1}
are straight lines and the peak feature at
$\theta \approx 90^\circ$ occurs at $90^\circ$ for all azimuthal
angles.

A possible explanation is that the true
interlayer spacing (as perceived by the quasiparticles)
is double the unit-cell height, or that there is a modulation
of the lattice in the crystallographic ${\bf b}$ direction;
this would result in an effective interlayer
spacing of $b^\prime =2\times 18.4$~\AA~$\approx 36.8$~\AA. Such a
modulation of the lattice could occur in the event of a
charge-density-wave (CDW) or spin-density-wave (SDW) instability;
however, there is as yet no other evidence for the presence
of such a groundstate
({\it c.f.} numerous other charge-transfer salts in
which CDWs or SDWs cause extensive modification
of the quantum-oscillation spectrum~\cite{review,eva}).
If such a modulation exists, it is too weak
to be picked up by a careful X-ray study at 115~K (Bruker-AXS
SMART6000 CCD, complete sphere of data, sixty-second frames, 0.3~degree
scans)~\cite{montxray}.  However,
this does not rule out the possibility that doubling occurs
at a lower temperature~\cite{montxray}.

Whatever
the mechanism, a doubling of $b^\prime$
(see Equation~\ref{amro1}) would result in the Fermi-surface parameters
$k_x=2.43 \pm 0.04~$nm$^{-1}$,
$k_y=0.815\pm 0.005~$nm$^{-1}$ and $\xi_{\rm inc}= 19 \pm 5^\circ$,
yielding an ellipse  area
corresponding to a Shubnikov-de Haas frequency of
$652\pm 15$~T, in very good agreement with $F_{\alpha}=650 \pm 5$~T
derived in Section 3.2.
Figure~\ref{locust}(b) shows an elliptical-cross-section
Q2D pocket of this size and
orientation alongside the most recent calculation
of the Fermi surface~\cite{whangbo}, based
on structural studies carried out at 17~K.
The Q2D pocket measured experimentally is smaller
and somewhat more
elongated than that suggested by the calculations;
it occupies $\approx 16$~\% of the Brillouin zone,
whereas the Q2D pocket of the calculation is 28~\% of the
Brillouin-zone area.
However, such discrepancies between calculation
and experiment are not without precedent in
crystalline organic metals~\cite{house,msamro,eva}.

\subsection{Estimate of the interplane transfer integral}
\label{squit}
We now turn to the small peak in the magnetoresistance
component $\rho_{zz}$ observed at $\theta =90^{\circ}$
in Figure~\ref{amro}.
Thus far, we have treated only the Fermi-surface
cross-section in the ${\bf a}^*{\bf c}^*$ plane
(see Figure~\ref{locust}(b)).
However, a small, but finite interlayer transfer
integral will lead to a warping of the Fermi surface in
the interlayer ${\bf b}^*$ direction (see Section~2 of
Reference~\cite{review}).
When the magnetic field is almost exactly in the plane of the warping,
a few closed orbits become possible
on the warped sections of the Fermi surface
(e.g. on the ``bellies'' of the warped Q2D Fermi
cylinders)~\cite{mck,hanasaki,goddard}.
For certain orientations of an in-plane field,
closed orbits will be possible on both the Q1D sheets
and Q2D cylinders of a Fermi surface such as that shown in
Figure~\ref{locust}(b); at other orientations of an in-plane field,
only the Q2D cylinders will be able to support closed orbits
on their bellies.
Such orbits are very effective at averaging
the interplane velocity component $v_z$,
and hence lead to a peak in
$\rho_{zz}$~\cite{hanasaki,russian,goddard}.

As the magnetic field is tilted away from the
in-plane direction ($\theta = 90^{\circ}$),
the closed orbits will cease
to be possible when $\theta=90^{\circ} \pm \Delta$,
where the angle $\Delta$ is given by
\begin{equation}
\Delta ({\rm in~radians})~\approx \frac{v_{\perp}}{v_{||}},
\label{vvvvv}
\end{equation}
where $v_{\perp}$ is the maximum interlayer quasiparticle velocity
and $v_{||}$ is the intralayer component of the quasiparticle velocity
in the plane of rotation of the magnetic field~\cite{hanasaki,russian,goddard}.
If the quasiparticle dispersion $E(k_b)$ in the interlayer
direction is assumed to follow a
simple tight-binding model, $E(k_b)=-2t_{\perp}\cos(k_bb)$~\cite{ashcroft},
where $t_{\perp}$ is the interlayer transfer integral,
then
\begin{equation}
v_{\perp}=2t_{\perp}b/\hbar,
\label{tvtv}
\end{equation}
where we have used the relationship
$\hbar {\bf v} =\nabla_{\bf k}E({\bf k})$~\cite{ashcroft}
to obtain $v_{\perp}$ from $E(k_b)$~\cite{goddard}.
Equations~\ref{vvvvv} and \ref{tvtv} therefore
show that there is a direct proportionality between $2\Delta$,
the angular width of the peak in $\rho_{zz}$, and the
interlayer transfer integral, $t_{\perp}$.

\begin{figure}[htbp]
\centering
\includegraphics[height=12cm]{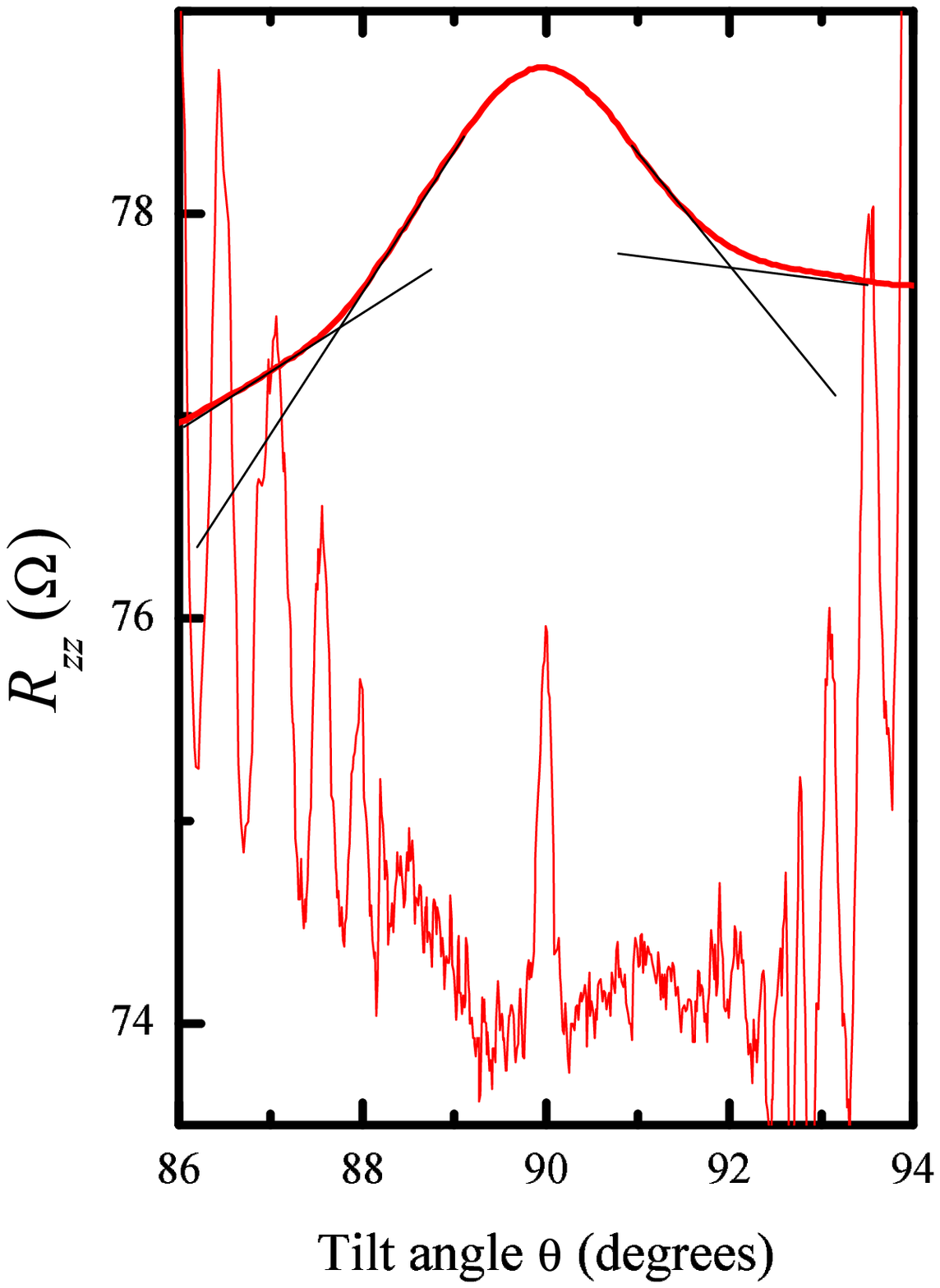}
\vspace{-0.4in}
\caption{Peaks in the interplane resistance $R_{zz}$
(proportional to $\rho_{zz}$) close to $\theta =90^{\circ}$ in
\lbets ~(thick line; $T=1.4$~K, $B=30$~T; present work) and
\cuscn ~(fine line; $T=520$~mK, $B=42$~T; Reference~\cite{goddard}).
In the case of \cuscn ,~the rapid oscillations at the edges of the
figure are angle-dependent magnetoresistance oscillations (AMROs).
The fine lines superimposed on the \lbets ~data show
how the full width of the peak is defined.}
\label{amro1a}
\end{figure}
Figure~\ref{amro1a} shows how the angular width
$2\Delta$ of the peak in $\rho_{zz}$
is defined; the limits of the peak are defined by the intersections
of extrapolations of the background magnetoresistance and the
edges of the peak~\cite{footies}. Similar data for
\cuscn ~from Reference~\cite{goddard} are plotted for comparison;
note how the peak at $90^{\circ}$ is much narrower.
Given the similarity of their intralayer Fermi-surface
properties (see Table~1), this comparison
immediately suggests a much smaller $t_{\perp}$ in \cuscn
~than in \lbets .

In order to obtain a quantitative estimate of $t_{\perp}$,
it is necessary to use reliable values of
$v_{||}$~\cite{hanasaki,goddard}.
Figure~\ref{locust}(b)
suggests that when the in-plane magnetic field is
close to the ${\bf a}^*$ direction, it is
likely that {\it both} Q1D and Q2D Fermi-surface sections will
be able to support closed orbits;
conversely, when the magnetic field is well away
from this orientation, only the Q2D cylinder
will be able to support closed orbits.
As most of the accurate information deduced from the
experiments in the previous Sections concerns
the Q2D Fermi-surface section,
we concentrate on the range of $\phi$ over which
it alone will determine the peak at $\theta=90^{\circ}$.

We assume an effective-mass-tensor
approximation for the in-plane motion on the Q2D Fermi-surface
section~\cite{review};
\begin{equation}
E=\frac{\hbar^2 k_x^2}{2m_1} +\frac{\hbar^2 k_y^2}{2m_2}
\label{hose}
\end{equation}
Here $m_1$ and $m_2$ are effective masses {\it for linear motion}
in the $x$ and $y$ directions. The cyclotron effective mass
in such an approximation is $m^*=(m_1m_2)^{1/2}$~\cite{ashcroft}.
Using the lengths of the axes of the elliptical cross-section
of the Q2D Fermi-surface section derived from the AMROs
(2.43~nm$^{-1}$ and 0.815~nm$^{-1}$; see previous section- we have divided
by 2 to account for apparent doubling of the unit-cell height)
and $m^*=3.6~m_{\rm e}$ (Table~\ref{kappatab}), we obtain
$\frac{m_1}{m_2}=\left(\frac{2.43}{0.815} \right)^2$
and
$m_1m_2=(3.6~m_{\rm e})^2=12.96~m_{\rm e}^2$,
yielding
$m_1=10.73~m_{\rm e}$,
$m_2=1.207~m_{\rm e}$ and $E_{\rm F}\approx 20.95$~meV
({\it c.f} $m_1=10.59~m_{\rm e}$,
$m_2=1.177~m_{\rm e}$ and $E_{\rm F}\approx 18.4$~meV
in \cuscn ~\cite{goddard}).
Hence, using $\hbar{\bf v}=\nabla_{\bf k}E({\bf k})$~\cite{ashcroft},
and constraining $E=E_{\rm F}$,
the velocities $v_{||}(\phi)$ may be derived
for the Q2D Fermi-surface section.

\begin{figure}[htbp]
\centering
\includegraphics[height=8cm]{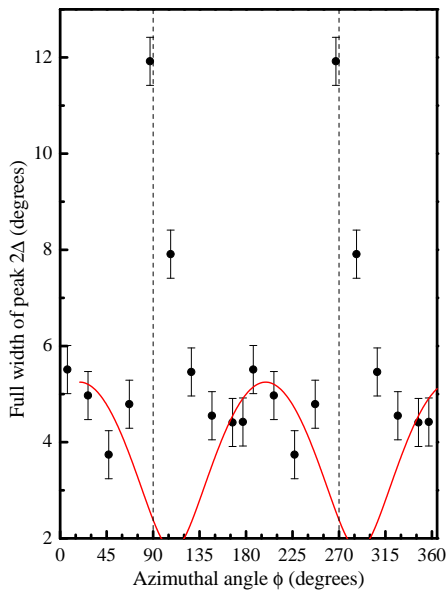}
\caption{Full widths of peaks in $\rho_{zz}$
close to $\theta =90^{\circ}$ for \lbets ~(points)
plotted as a function of $\phi$.
The curve is given by Equations~\ref{vvvvv}, \ref{tvtv}
and \ref{hose}, with $t_{\perp}=0.21$~meV.
Note that we have chosen to fit data for ranges of $\phi$ at which
the Q2D pocket is expected to be the {\it sole} provider of
closed orbits for in-plane magnetic fields;
{\it i.e.} we avoided field orientations close
to $-{\bf a}^*$ ($\phi=90^{\circ}$) and ${\bf a}^*$ ($\phi=270^{\circ}$)
(dashed lines) at which the Q1D sheets
might also be expected to provide closed
orbits in an in-plane field (see Figure~\ref{locust}(b))
}
\label{amro2}
\end{figure}

The widths $2 \Delta$ derived from data such as those in
Figure~\ref{amro1a} are plotted as a function of
$\phi$ in Figure~\ref{amro2}.
The Figure also shows the prediction of Equation~\ref{vvvvv}
using $v_{||}$ derived from Equation~\ref{hose};
the only fit parameter is $t_{\perp}$ (see Equation~\ref{tvtv}).
We have chosen to fit data for ranges of $\phi$ at which
the Q2D pocket is expected to be the {\it sole} provider of
closed orbits for in-plane magnetic fields;
{\it i.e.} we avoided field orientations close
to $-{\bf a}^*$ ($\phi=90^{\circ}$) and ${\bf a}^*$ ($\phi=270^{\circ}$)
(dashed lines) at which the Q1D sheets
might also be expected to provide closed
orbits in an in-plane field (see Figure~\ref{locust}(b)).
(Note that the experimental data show a strong peak
close to $\phi=90^{\circ}$ and $270^{\circ}$ (instead of the minimum
predicted by the model)
suggesting that the Q1D sections are indeed the dominant cause of
the peak in $\rho_{zz}$ at these orientations.)
The fit yields $v_{\perp}\approx 1200$~ms$^{-1}$,
so that $t_{\perp} \approx 0.21$~meV~\cite{footnote22}.
A similar procedure has been carried out for \cuscn ,
giving an interplane transfer integral of $t_{\perp} \approx 0.04$~meV~\cite{goddard}.
Therefore, the interplane transfer integral in \lbets
~is a factor $\sim 5$ bigger than that in \cuscn .

\subsection{Summary.}
In summary, the magnetoresistance measurements indicate that
the Fermi surface of
\lbets ~bears a strong resemblance to
that of \cuscn ~within the highly-conducting Q2D planes.
Moreover, the effective masses in the two salts are almost identical
and the renormalising interactions are probably of a similar strength.

The width of the peak in the
magnetoresistance close to $\theta =90^{\circ}$
suggests that the interplane transfer integral in
\lbets ~is approximately five times bigger than that
in \cuscn . This implies that
\lbets ~is a less two-dimensional
material.
\section{Penetration-depth measurements.}
\subsection{Experimental details.}
\label{penetrate}
In the current experiments, the penetration depth was inferred by
placing the superconducting sample in a small coil which
is the inductive element of a resonant tank circuit~\cite{tdo,janeloff}.
The exclusion of flux from the sample, and hence the coil,
decreases the inductance of the circuit;
consequently the resonant angular frequency, $\omega = 1/\sqrt{LC}$, will
increase.
The well-known properties of inductors~\cite{bleaney} lead
to $\frac{\Delta A_{\phi}}{A_{C}}=\frac{\Delta L}{L}$,
where ${\Delta A_{\phi}}$ is the change in flux area, $A_{\rm C}$ is the area
of the measurement coil, $\Delta L$ is the change in
inductance and $L$ is the total inductance.
For small changes in
inductance,  $\frac{\Delta L}{L_{0}}=2\frac{\Delta f}{f_{0}}$,
where $\Delta f$ is the change in resonant frequency, $f_{0}$ is the
initial resonant frequency, and $L_{0}$ is the initial inductance.
Through simple geometrical
relations it can be shown that~\cite{mielkethesis}
\begin{equation}
\frac{\Delta A_{\phi}}{A_{C}}=
\frac{2r_{s}\Delta \lambda - \Delta \lambda^{2}}{R^{2}},
\end{equation}
where $\Delta \lambda$ is the change in penetration depth, $R$
is the  effective radius of the coil, $r_{s}$ is the effective sample radius.
Simple estimates for the sample sizes used in the current experiments
show that the second-order term is
negligible~\cite{mielkethesis}, so that
\begin{equation}
\Delta\lambda=\frac{R^{2}}{r_{s}}\frac{\Delta f}{f_{0}}.
\label{maddog}
\end{equation}

The \lbets ~samples used in the current study
have a needle-like geometry~\cite{grow2}; a typical example
had approximate dimensions $2 \times  0.170 \times 0.084$~mm$^3$.
In order to maximize the filling factor and cross sectional
area, the coil was made rectangular, with an effective area
of $1.34$~mm$^2$; for the sample mentioned above,
the effective sample radius is half the shortest dimension
(0.042~mm) (note that the long axis of the crystal is perpendicular
to the coil axis).
Calibration was achieved by placing a
spherical superconducting sphere of known size in the
coil~\cite{carrington}.
The sample was orientated in the coil
so that the oscillating magnetic field was
parallel to the crystallographic ${\bf b}^*$ direction, 
{\it i.e.}~perpendicular to the highly-conducting planes~\cite{growthdetails}.
The tank-circuit capacitance was provided by a
30~pF mica capacitor, and the circuit was driven
at $f \approx 25$~MHz by a tunnel-diode oscillator~\cite{tdo,footnote1}.
The sample and coil were placed in a $^3$He cryostat or dilution refrigerator.
Quasistatic magnetic
fields of up to 30~T were applied parallel to ${\bf b}^*$. 

\begin{figure}[htbp]
\centering
\includegraphics[height=8cm]{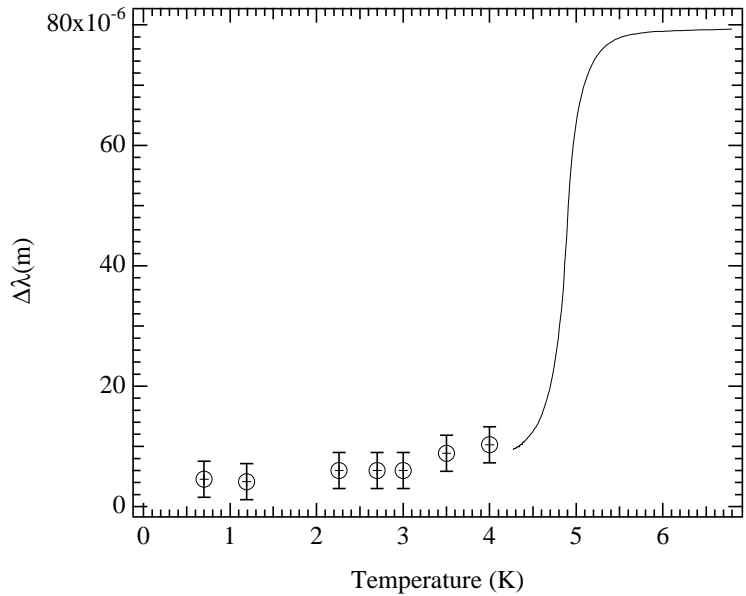}
\caption{Change in penetration depth versus temperature at zero field in \lbets .
Note that the cryostat used could only provide controlled, slow sweeps
of temperature down to 4.2~K (data shown as continuous curve).
Below this, data are recorded at fixed, stable temperatures (points).
The error bars on the points give typical uncertainties,
valid across the whole temperature range shown.}
\label{lamvt}
 \end{figure}

Figure~\ref{lamvt} shows $\Delta \lambda$ of a \lbets ~crystal
(deduced from the frequency of the tank circuit
using Equation~\ref{maddog}) as a
function of temperature.
The sharp rise in penetration between
4.5~K and 5.1~K gives a very clear indication
of the superconducting to normal transition; above this
temperature $\Delta \lambda \approx 80~\mu$m
(i.e. comparable to the shortest dimension of the sample),
indicating that the radiofrequency fields penetrate the whole crystal
once it is in the normal state.
This is in agreement with estimates of the low-temperature normal-state conductivity for
\lbets~\cite{calculation}, which lead one to expect an in-plane skin depth $\delta \sim
100~\mu$m~\cite{schrama,hill}.
\subsection{Measurements at low magnetic fields: determination of pinning parameters}
Wu and Sridhar~\cite{wu}
have treated the repulsively interacting flux lines in a
type-II superconductor as periodic,
damped harmonic-oscillator potentials modulated by a rf
field.  The physical
justification of their model is that the repulsive interaction of the
fluxoids causes the Abrikosov lattice~\cite{abrikosov,tinkham}
to resist higher flux densities in a
manner analogous to the way in which a two dimensional
network of springs
resists compression. In such a model,
the Labusch pinning potential parameter $\alpha$
corresponds to the restoring force on fluxoids displaced
slightly by the current density $J$ induced
by the radiofrequency field~\cite{wu}:
$\eta \frac{{\rm d}x}{{\rm d}t}+ \alpha x =\phi_0 J$.
Here $x$ is the fluxoid displacement,
$\eta$ is a damping parameter and $\phi_0$ is the flux quantum~\cite{wu}.
In other words,
a small perturbation displaces
the fluxoid from its equilibrium position against the restoring force
provided by the repulsion from neighbouring fluxoids
and the pinning potential.

\begin{figure}[htbp]
\centering
\includegraphics[height=12cm]{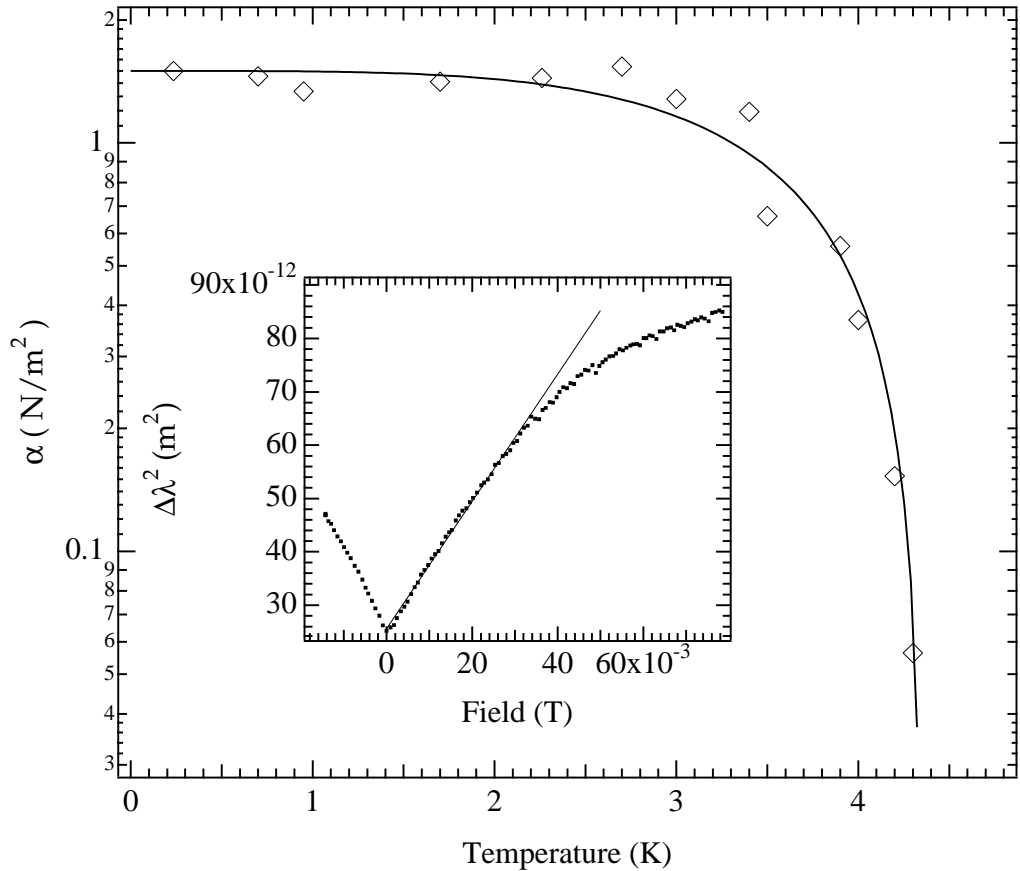}
\caption{Inset: the determination of the Labush parameter $\alpha(T)$ by
a straight line fit to low-field $\Delta\lambda^{2}$ vs. $\mu_{0}H$ data
($T=700$~mK). Main figure: experimental
values of $\alpha$ versus temperature (points); the curve
is a fit to the two-fluid model expression.}
\label{alpha}
\end{figure}

In the limit of small magnetic field ($H \ll B_{\rm c2}$),
the damping due to fluxoid viscous drag~\cite{wu,bs}
may be neglected, leading to a linear relationship
between changes in
the square of the penetration depth $\Delta \lambda^2$
and the magnetic induction $B$ inside the sample~\cite{wu},
\begin{equation}
\Delta \lambda^2=\frac{\phi_0}{\mu_0 \alpha(T)}B(H).
\label{gunge}
\end{equation}
Figure~\ref{alpha} (inset) shows the field dependence of $\Delta\lambda^{2}$
at a temperature of $T=700$~mK.
The sample is well inside the mixed state for fields above $\mu_0H \approx 0.01$~T,
so that $B=\mu_0H$; hence, the linear dependence predicted by
Equation~\ref{gunge}
fits the data well, yielding a Labusch
parameter of $\alpha(700~{\rm mK}) \approx 1.4$~Nm$^{-2}$.

Data similar to those in the inset of Figure~\ref{alpha} were acquired at a range
of temperatures; the resulting values of $\alpha$ are plotted in
Figure~\ref{alpha} (main figure) as a function of temperature.
The solid line fit through the data in
Figure~\ref{alpha} is the predicted temperature dependence from the
two-fluid Gorter-Casmir~\cite{tinkham,gorter} model,
\begin{equation}
    \alpha \propto \left[1-\left( \frac{T}{T_{\rm c}} \right)^{4}\right].
\end{equation}

Figure~\ref{alpha} also highlights some of the differences between
organic superconductors and the ``high-$T_{\rm c}$'' cuprates.
As $T \rightarrow 0$, the Labusch parameter in \lbets ~
tends to  $1.5$~Nm$^{-2}$, almost
four orders of magnitude smaller than that in YBa$_2$Cu$_3$O$_7$~\cite{wu}.
Moreover, whereas the two-fluid model is able to describe the
\lbets ~data to a fair degree of accuracy, the pinning-force
parameter in YBa$_2$Cu$_3$O$_7$ has been
shown to follow the temperature dependence $(1 - (T/T_{\rm c})^{2})^{2}$~\cite{wu}.
\subsection{Measurements at high magnetic fields:
deviation from Campbell penetration-depth behaviour and
upper critical field.}
\label{bleou}
\begin{figure}[htbp]
\centering
\includegraphics[height=8cm]{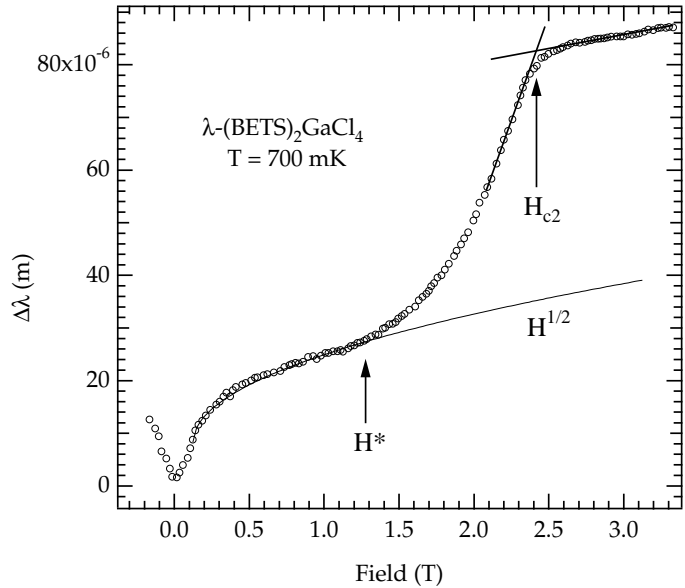}
\caption{The change in penetration depth signal vs.  field for \lbets
at 700 mK. The deviation from the Campbell penetration depth behavior
is indicated as $H^{\ast}$ while the saturation of $\Delta\lambda$
signal indicates $H_{\rm c2}$, determined from the straight line intercepts.}
    \label{fig:dlvsB}
\end{figure}
Figure~\ref{fig:dlvsB} shows the measured change in penetration
depth $\Delta \lambda$ for \lbets ~ at
intermediate fields and a temperature of 700~mK.
$\Delta \lambda$ varies approximately
as $\sqrt{ H}$, as expected in the
Campbell scenario for fluxoid motion~\cite{wu,campbell}.
However, above a field which we label $\mu_0 H^*$,
$\Delta \lambda$ deviates from the $\sqrt{ H}$
dependence; this implies that
the periodic harmonic well models~\cite{wu,cnc} are
no longer applicable.
At fields above $\mu_0H^*$, $\Delta \lambda$ follows
the approximate field dependence $H^2$, until
the penetration depth saturates.

The identification of the upper critical field
in organic superconductors from conductivity data
has been the subject of
considerable debate~\cite{hcpapers,belin};
the transition is intrinsically broad, and phenomena
such as a pronounced ``hump'' in the resistivity
and negative magnetoresistance are observed close
to $H_{\rm c2}$~\cite{review}. (Note that
similar complications also afflict the ``High-$T_{\rm c}$''
cuprates~\cite{finnemore,boeb}.)
Recently, a concensus has emerged whereby most
of the transition region between zero resistance
and normal-state magnetoresistance is regarded as
a property of the mixed phase (see References~\cite{janeloff,belin}
and references therein); $H_{\rm c2}$
is then defined as the intersection of the
extrapolations of the transition region and the normal-state
magnetoresistance~\cite{janeloff}.

In penetration depth measurements, the broadening is less
severe since the measurement is not dependent on a macroscopic net
current flow across the sample~\cite{wu,belin}.
The pinned fluxoids probed by rf
fields do not experience as large an electric field gradient and hence
the dissipation associated with the normal core is reduced~\cite{wu,belin}.
Nevertheless, the transition is still somewhat broadened,
and so we follow the same spirit of the convention
used in resistivity studies~\cite{janeloff} and
the GHz penetration depth studies of Reference~\cite{belin},
defining $H_{\rm c2}$ as
the intersection of extrapolations of the
penetration-depth curves below and above the
point at which the saturation occurs~(see Figure~\ref{fig:dlvsB}).
We can be confident that the saturated behaviour
is characteristic of the normal state, as it continues
up to at least 30~T without further features.
This strongly suggests that the whole of the
sample is penetrated by the rf fields in the
normal state, as suggested in Section~\ref{penetrate}.

\begin{figure}[htbp]
\centering
\includegraphics[height=10cm]{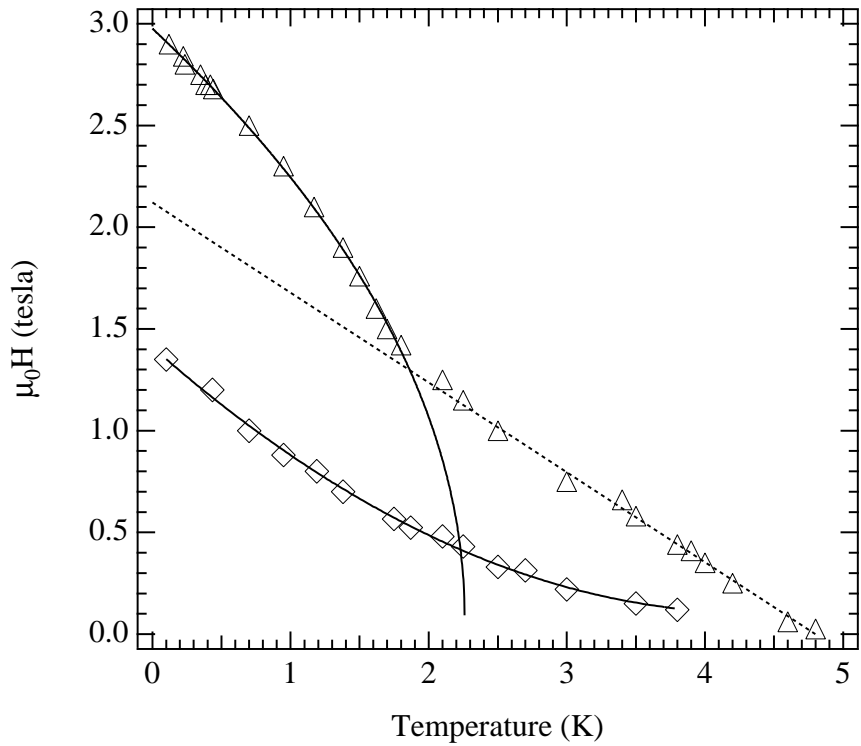}
\caption{Notional phase diagram of \lbets ~showing the upper critical
field (triangles) and $H^{\ast}$, marking the inflection points
in the penetration depth (diamonds).
The quasistatic field is
applied parallel to ${\bf b}^*$, {\it i.e.} perpendicular
to the quasi-two-dimensional planes of the sample.
Points from four different samples are shown, often overlaying each other.
The upper solid curve is $H_{\rm c2} \propto (T^*-T)^{1/2}$;
the dashed curve is $H_{\rm c2} \propto (T_{\rm c2}-T)$.
The lower solid curve is a fit of the two-fluid model
expression for the flux-line lattice melting.
}
\label{fig:ht}
\end{figure}

The values of $H_{\rm c2}$ and $H^*$ deduced from
the pentration depth measurements are shown as a function
of temperature in Figure~\ref{fig:ht}.
Four different \lbets ~samples, taken from four
separate growth batches, were used in the study.
There were negligible differences in their behaviour,
and the data from the four samples in Figure~\ref{fig:ht}
overlie each other, suggesting that the characteristic fields measured
are intrinsic properties of \lbets .
We shall return to the temperature dependence of $H_{\rm c2}$
in a later section.

\subsection{Flux-lattice melting at $H^*$.}
We now turn to the change in behaviour which occurs
at the field $H^*$ (see Figure~\ref{fig:dlvsB}).
We attribute the change at $H^*$ to flux-line lattice
melting, as $H^*$ follows the $(T_{\rm c}-T)^2$
dependence expected from the Gorter-Casimir
two-fluid model~\cite{gorter}.

Additional support for this attribution comes from considering
microscopic models of the melting process.
Houghton {\it et al.}~\cite{houghton} have considered
the elastic moduli of the flux-line lattice and
proposed that melting occurs when the mean thermal flux-line
displacement $d(T)$ is a substantial fraction of
the Abrikosov lattice parameter $\ell=(2 \phi_0/\sqrt{3} B)^{1/2}$, {\it i.e.},
$d(T) \approx c_{\rm L} \ell$.
Here $c_{\rm L}$ is the Lindemann parameter~\cite{saturationbombing},
a function used very generally in the description of solid-liquid transitions;
typically $c_{\rm L} \sim 0.1-0.2$.
The explicit expression
for $d(T)$ is
\begin{equation}
    d(T)=\sqrt{\frac{1}{2\pi}\left(\frac{G_{i}}{\gamma^{2}}\right)^{1/2}\frac{t}
    {(1-t)^{1/2}}\frac{b}{(1-b)}\left(\frac{4(\sqrt{2}-1)}{(1-b)^{1/2}}+1\right)
    \ell^{2}}
    \label{eq:d}
\end{equation}
where $t=T/T_{\rm c}$ and $b=B/B_{\rm c2}$.
The parameter
$G_{i}=(16\pi^{3}\kappa^{4}(k_{\rm B}T)^{2})/(\phi_{0}^{3}H_{c2}(0))$
describes the importance of fluctuations in a given
system; $\gamma$ is the anisotropy term defined by
\begin{equation}
    \gamma\equiv
    \frac{\xi_{xy}}{\xi_{z}},
\label{xixixi}
\end{equation}
with $\xi_{xy}$ and $\xi_z$ being the in-plane and interplane
coherence lengths respectively.
$\kappa$ is another Ginzburg-Landau parameter giving the ratio of
penetration depth to the coherence length for a particular orientation~\cite{tinkham,houghton}.
In this case we require $\kappa=\frac{\lambda}{\xi_{\perp}}$.
Using Ginzberg-Landau theory in the limit $T \rightarrow 0$~\cite{guelph}
yields $\xi_{\perp}\approx$ 1.4~nm and $\lambda \approx$150~nm for \lbets ,
resulting in $\kappa \approx 107$.

In this model, the melting
field at each temperature can be interpreted as the
point at which the product of $c_{\rm L}$ and the
lattice spacing is roughly equal to the average flux line
displacement. Thus, an increasing field reduces
the average inter-vortex spacing, thereby
facilitating melting~\cite{houghton}.
Substituting $\mu_0H^* \approx 1$~T at $T=700$~mK and $\kappa =107$
into the above equations yields a flux-line displacement
of approximately 6.0~nm,
roughly 12$\%$ of the Abrikosov lattice spacing.
This implies that
$c_{\rm L} \approx 0.12$,
a value entirely typical of a solid-liquid transition~\cite{saturationbombing}.

In isotropic superconducting systems, the melting of the flux-line lattice
occurs so close to $H_{\rm c2}$
as to be indistinguishable from it~\cite{tinkham}.
However, in materials such as \lbets ~(and in \cuscn ; see References~\cite{sasaki,mola}),
the large anisotropy of the superconducting
properties permits the melting line to be observed over
extended regions of the $H-T$
phase diagram, well clear of $H_{\rm c2}$.
\section{Discussion: comparison of \lbets ~with \cuscn ; dimensional cross-over.}
\label{discussion}
Having seen in Section~\ref{bands} that the Fermi surfaces
of \lbets ~and \cuscn ~bear some striking similarities
within the Q2D planes but have interplane transfer integrals
differing by a factor $\sim 5$, it is interesting to compare
their superconducting properties.

\begin{figure}[htbp]
\centering
\includegraphics[height=10cm]{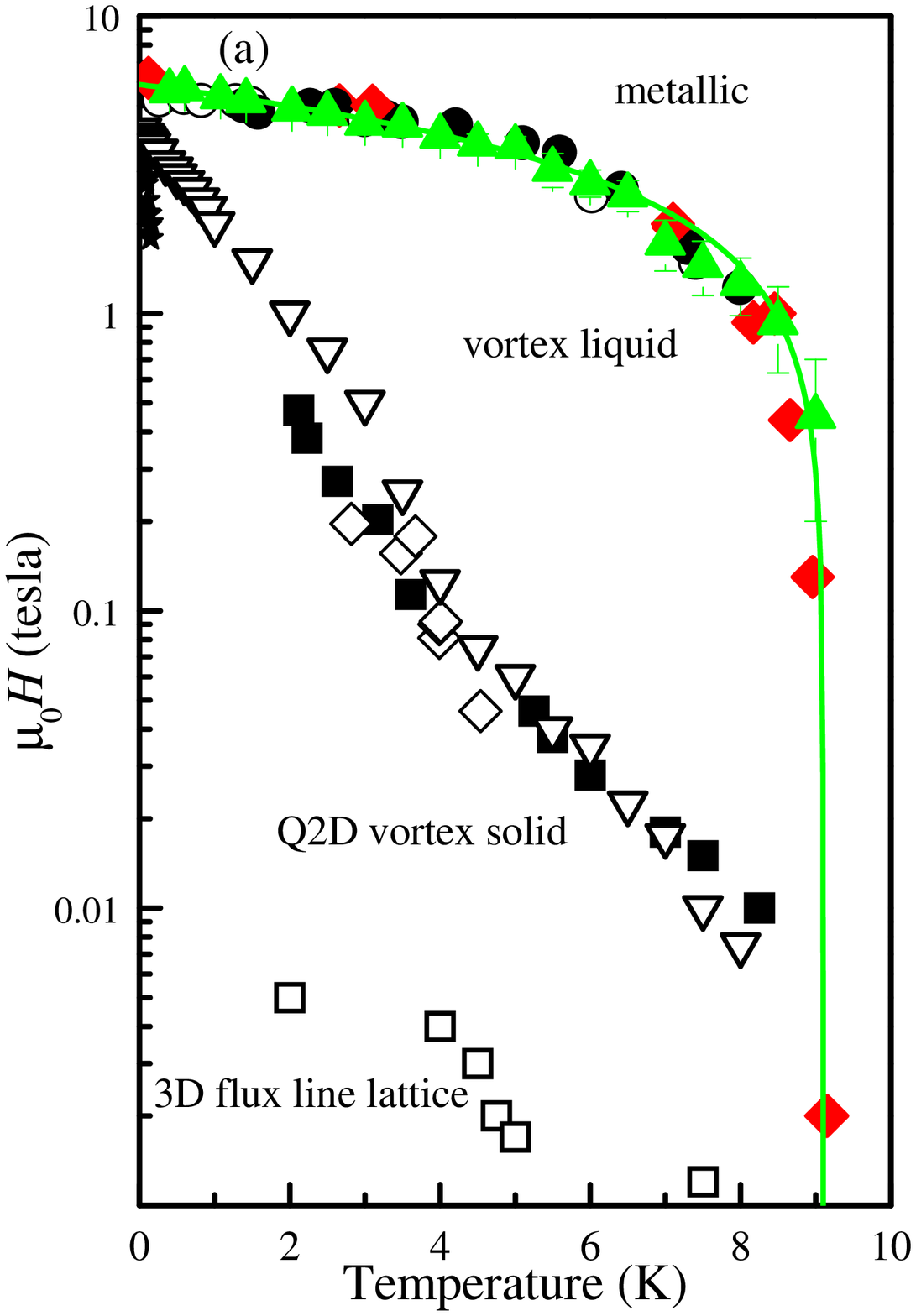}
\includegraphics[height=10cm]{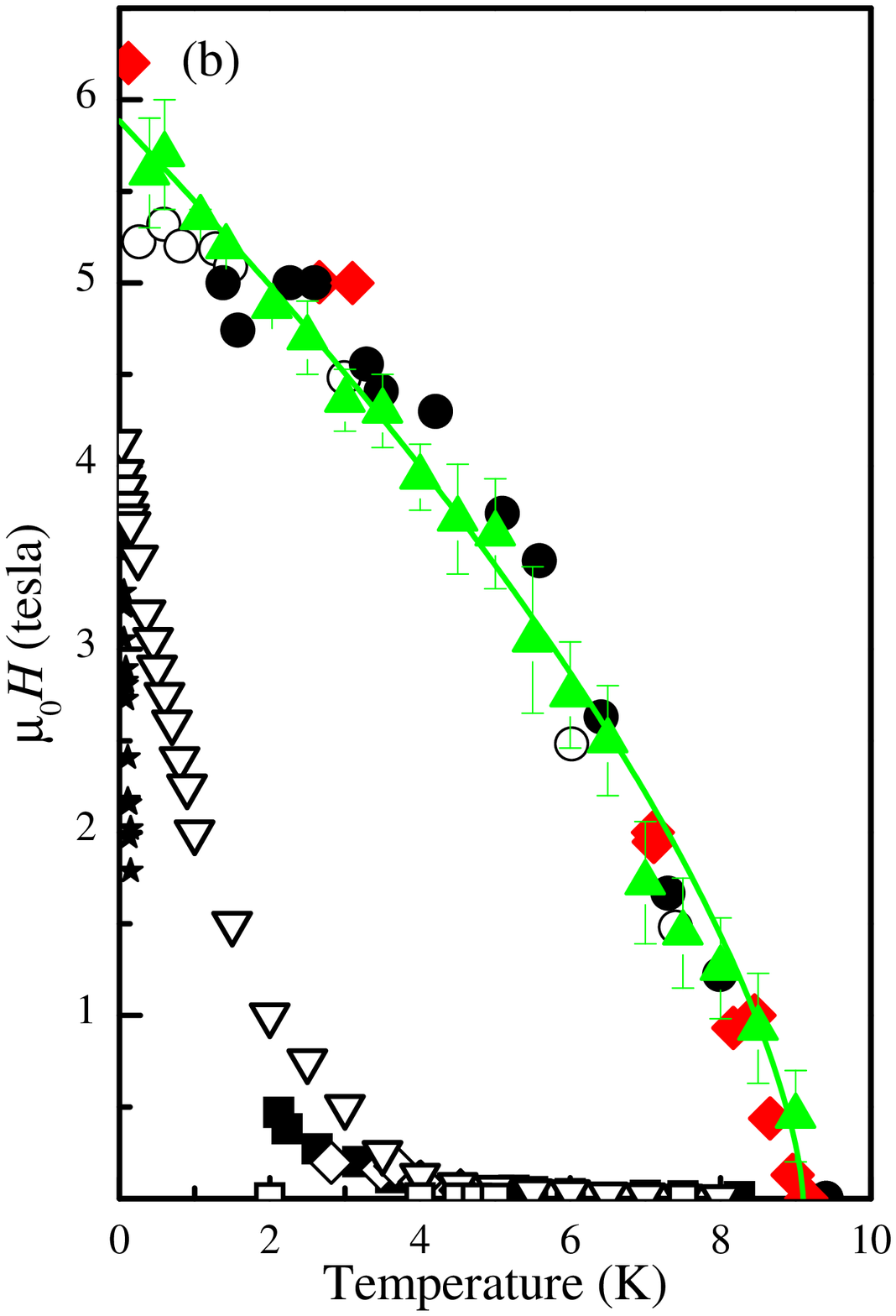}
\caption{Critical fields in \cuscn , plotted on
logarithmic (a) and linear (b)
field scales. The data for $H_{\rm c2}$ comprise filled triangles
(MHz penetration data obatined using the apparatus described
in the current paper~\cite{jscm})
filled circles (microwave penetration studies; Reference~\cite{belin};
errors in $\mu_0 H_{\rm c2}$ values typically $\pm 0.5$~T),
open circles (thermal conductivity data of Reference~\cite{belin};
errors not given) and shaded diamonds (magnetisation data;
the lowest temperature point was determined by examining the
attenuation of de Haas-van Alphen oscillations~\cite{sasaki} and
the higher-temperature points are from the scaling studies
in Reference~\cite{lang}).
The solid curve is proportional to $(T_{\rm c}-T)^{2/3}$,
with $T_{\rm c} =9.1$~K.
The triangles are the irreversibility field from magnetisation~\cite{sasaki};
the filled squares and stars represent 2D melting from magnetometry
and GHz studies~\cite{mola} (see also the NMR
data of Reference~\cite{nmrnmr}). The hollow squares are from
muon-spin rotation~\cite{leepratt} and denote the 3D-2D
transition.
}
\label{cuncs}
\end{figure}

The work of Belin {\it et al.}~\cite{belin}
has shown that conventional
resistivity measurements can yield unrepresentative values
for the upper critical field of \cuscn ~(see also \cite{boeb,graebner});
the difficulties
result from the dissipative mechanisms mentioned
in Section~\ref{bleou}, which act to broaden the resistive
transition~\cite{review,hcpapers}.
Thermal conductivity, magnetisation and penetration depth measurements
seem to be less susceptible to these problems and give a
better reflection of the true $H_{\rm c2}$~\cite{belin}.
We have therefore compiled the $H_{\rm c2}(T)$ plot in Figure~\ref{cuncs}
using available thermal conductivity~\cite{belin},
magnetisation~\cite{sasaki,lang} and MHz~\cite{jscm}
and microwave ($12-25$~GHz)~\cite{belin} penetration
measurements.
There is some scatter amongst the data from
different measurements,
but all suggest that $(\partial H_{\rm c2}/\partial T)$
increases in magnitude as $T$ increases, and
in fact the power law $H_{\rm c2} \propto (T_{\rm c}-T)^{2/3}$ is
quite successful in describing the data (Figure~\ref{cuncs}).

A material made up of weakly-coupled superconducting
planes may transform from
a three-dimensional system to what is in effect
a series of two-dimensional superconductors
as the interlayer coherence length decreases
with decreasing temperature~\cite{ld}.
The dimensional crossover occurs when the interplane coherence length
$\xi_z$ becomes shorter than the interplane spacing of the
quasi-two-dimensional layers.
Muon-spin rotation studies~\cite{leepratt} have shown that this
transition occurs in \cuscn ~at magnetic fields
$\sim 7$~mT, {\it i.e.} three orders of magnitude
smaller than typical values of $H_{\rm c2}$ (Figure~\ref{cuncs}).
This strongly suggests that the variation of $H_{\rm c2}$
shown in Figure~\ref{cuncs} is typical of a
quasi-two-dimensional
superconductor consisting of
weakly-coupled layers~\cite{sasaki,mola,graybeal}.

Figure~\ref{fig:ht} shows that the upper critical field of \lbets
~(conducting planes perpendicular to the applied magnetic field) has a
linear region $H_{\rm c2} \propto (T_{\rm c}-T)$
that spans from $T_{c}$ to approximately 1.9~K. Below
1.9~K, a definite change in the slope of the upper critical field
occurs, and $H_{\rm c2}$ begins to follow the power
law $H_{\rm c2} \propto (T^*-T)^{\zeta}$, with $T^*$
a fit parameter; powers
$\zeta$ in the range $0.5-0.7$ provide an adequate fit
to the data.

The behaviour of $H_{\rm c2}$ in \lbets ~at temperatures
below 1.9~K is therefore very similar to that of $H_{\rm c2}$
in \cuscn ~over the whole temperature range shown in
Figure~\ref{cuncs}, and is thus characteristic
of a two-dimensional
superconductor with weakly-coupled layers~\cite{mola,graybeal}.
On the other hand, the linear variation
of $H_{\rm c2}$ in \lbets ~at higher temperatures
follows the expectations of Ginzberg-Landau theory
for three-dimensional superconductors~\cite{tinkham,parks}.
We therefore attribute the change in gradient at 1.9~K
to dimensional cross-over from quasi-two-dimensional (low temperatures)
to three dimensional (high temperatures).
As we shall now show, this is entirely consistent
with estimates of the interplane coherence length in \lbets .

The magnetic-field-orientation dependence of $B_{\rm c2}$~\cite{mielkethesis}
in \lbets ~is qualitatively similar
to the predictions
of the Ginzberg-Landau
anisotropic effective mass approximation~\cite{tinkham,naka,Morris,notethis}
\begin{equation}
B_{\rm c2}(\theta) = \frac{B_{\rm c2}(\theta = 0)}
{\sqrt{\cos^{2}(\theta) +  \gamma^{-2} \sin^{2}(\theta)}} ,
\label{gl}
\end{equation}
where $\theta$ is the angle between ${\bf b}^*$ and the applied magnetic field
and $\gamma$ has been defined in Equation~\ref{xixixi}.
Using Ginzberg-Landau theory~\cite{tinkham},
the in-plane coherence length
may be estimated from the upper critical field
when the magnetic field is parallel to ${\bf b}^*$:
\begin{equation}
H_{\rm c2}(\theta =0)=\frac{\phi_0}{2 \pi \xi^2_{xy}}.
\label{xixixixi}
\end{equation}

Fits of the $\theta$ dependence of $H_{\rm c2}$ at $T=1.75$~K~\cite{mielkenew,notethis}
(the temperature at which the change in gradient in
Figure~\ref{fig:ht} occurs) yield $\xi_z \approx 1.85$~nm,
almost identical with the interplane spacing, and
supporting our assertion that the change in gradient
in Figure~\ref{fig:ht} at 1.9~K is associated with
a dimensional crossover.

Dimensional crossovers with the magnetic field
applied perpendicular to the Q2D planes
have been observed
in artificial Q2D superconducting structures~\cite{white,perpnote}
and in organic superconductors such as \cuscn~\cite{leepratt,mola};
however, in the majority of these cases, the effect of the crossover
is observed at magnetic fields less than $H_{\rm c2}$.
\lbets ~is perhaps unique in providing the correct
anisotropy for the crossover to be observed in the
behaviour of $H_{\rm c2}(T)$.

In Section~\ref{squit} we demonstrated that the interplane
transfer integral in \lbets ~is a factor $\sim 5$ larger than
that in \cuscn . The greater ``three dimensionality'' of
the bandstructure of \lbets ~compared to
\cuscn  ~obviously manifests itself in the
superconducting behaviour (compare Figures~\ref{fig:ht}
and \ref{cuncs}); whereas \lbets ~exhibits
2D-3D dimensional crossover in its $H_{\rm c2}(T)$ behaviour,
$H_{\rm c2}(T)$ in \cuscn ~is entirely characteristic of a Q2D
superconductor.
\section{Summary}
In summary, we have measured the Fermi surface topology of the organic
superconductor \lbets
~using Shubnikov-de Haas and angle-dependent magnetoresistance oscillations.
The data show that the Fermi-surface topology of
\lbets ~is very similar indeed to that of the most heavily-studied
organic superconductor, \cuscn , except in one important
respect; the interplane transfer integral in \lbets
~is a factor $\sim 5$ larger than that in \cuscn .
The increased three-dimensionality of \lbets
~is manifested
in radiofrequency penetration-depth measurements,
which show a clear dimensional crossover in the behaviour
of $H_{\rm c2}$. The radiofrequency measurements have also
been used to extract the Labusch parameter determining
the fluxoid interactions as a function of temperature,
and to map the flux-lattice melting curve.

We have observed a discrepancy between the
angle-dependent magnetoresistance oscillation and Shubnikov-de Haas
data which suggests that the true unit cell height
at low temperatures is double that inferred
from X-ray studies. At present, this has not been detected by other techniques.

It is interesting to note that the anisotropies
and Ginzberg-Landau parameters of the organic superconductors
\lbets  ~(this work) and \cuscn ~\cite{leepratt,sasaki,mola}
span the typical values found in ``High $T_{\rm c}$'' cuprates
such as YBCO and BISCCO~\cite{tinkham}.
However, as the current work has shown, in contrast to the cuprates,
the Fermi-surface topologies and complete phase diagrams of
organic superconductors such as \lbets ~and \cuscn ~can be
mapped out in detail using accessible laboratory fields.
Moreover, details of the bandstructure in the organics
can be related directly to the superconducting properties.
The availability of a large number of organic superconductors
of varying dimensionality and bandstructure~\cite{review,tossnitza}
should potentially allow very stringent experimental tests of models
of superconductivity in layered materials to be carried out.
\section{Acknowledgements.}
This work is supported by the Department of Energy, the National
Science Foundation (NSF), the State of Florida and EPSRC (UK).
Acknowledgement is made to the donors of The Petroleum Research Fund,
administered by the ACS, for partial support of this research.
CHM thanks J.S. Brooks for very useful insights into
the radiofrequency measurement technique.
We are grateful to Stephen Hill and Monty Mola for their considerable
help in compiling
Figure~\ref{cuncs} and Kazumi Maki, Ross McKenzie and Stephen Blundell
for illuminating discussions.
Paul Goddard, Mike Whangbo and J.-H. Koo
are thanked for permission to use data and calculations
from References~\cite{whangbo,goddard} prior to publication.
Finally, we acknowledge the suggestions of one of the Referees,
which have helped to clarify several points.

\section{Appendix: a note on the radiofrequency response.}
In the current paper we have extracted parameters
which describe the superconducting state of
\lbets ~under the assumption that all of the apparent changes
in the penetration of the radiofrequency field
are due to the variation of $\lambda$.
It is therefore very important to assess whether this
assumption is valid.
Moreover, as little has been written in the
literature about the radiofrequency techniques employed,
it is useful to summarise the artefacts which can affect
the experimental data.
For future reference, we hope 
that it is also useful to provide
estimates of some of the paramters used in the
theory used to model experimental data~\cite{cnc}.

Coffey and Clem~\cite{cnc} have treated the behaviour of a
superconductor in a rf field over a broad frequency range. 
Using their approach, the contributions
to the penetration depth signal
from surface-impedance and skin-depth effects can be
evaluated as the temperature or applied magnetic field are varied. 
The model defines boundaries at which
the surface-impedance and skin-depth effects become
non-negligible; these are set by evaluating the flux creep factor~\cite{cnc}.

The flux creep factor $\varepsilon$ is determined by $\nu$,
the ratio
of the fluxoid barrier height $U_0$ to the 
typical thermal energy $k_{\rm B}T$,
$\nu=\frac{U_{0}}{2k_{\rm B}T}$.
The flux creep factor is then determined by: 
$\varepsilon=1/I_{0}(\nu)^{2}$,
where $I_{0}$ is a zeroth-order
modified Bessel function of the first kind~\cite{cnc}.
$\varepsilon$ parameterises the
degree to which thermal effects assist
the motion of fluxoids.

In the limit of large $\varepsilon$ ({\it i.e.} $\varepsilon \sim 1$),
thermal excitation causes the behaviour of
fluxoids to approach that of completely unpinned fluxoids.
In this case, the complex effective resistivity ($\tilde{\rho}_{v}(\omega)$)
becomes a contributing factor to the measured change in penetration
depth.  $\tilde{\rho}_{v}(\omega)$~\cite{cnc} is given by the
expression
\begin{equation}  
    \tilde{\rho}_{v}(\omega)=\frac{\varepsilon+(\omega\tau)^{2}
    +i(1-\varepsilon)\omega\tau}{1+(\omega\tau)^{2}}\rho_{f},
    \label{comp}
\end{equation}    
where $\rho_{f}=B\phi_{0}/\eta$ is the flux flow resistivity
and $\omega\tau$ represents the product of the measurement frequency
and the relaxation time of the normal-state quasiparticles.

An order-of-magnitude estimate of $U_0$ is given by considering
the energy at which the harmonic-oscillator potentials
of neighbouring fluxoids cross, yielding~\cite{wu}
$U_0 \approx \frac{2}{\pi^2}\alpha L^3$.
The charcteristic length $L$~\cite{wu} will be roughly
equal to the Abrikosov lattice spacing,
$L=\ell$,
so that the pinning well barrier height $U_0$
can be estimated using
\begin{equation}
		U_{0}=\frac{2 \alpha}{\pi^2}\left(\frac{2}{\sqrt{3}}
		\frac{\phi_{0}}{B}\right)^{3/2}.
	\label{U3}
\end{equation}

Equation~\ref{U3} shows that $U_0$ decreases as the field
$B$ increases, so that the
surface-impedance and skin-depth effects will be most prominent
at high magnetic fields.
Within the superconducting state, the highest field ({\it i.e.}
worst-case scenario) at which
we make quantitative deductions about vortex behaviour
is $\mu_0 H^*$, the field at which the penetration depth indicates a 
divergence from the Campbell regime.
Substituting the value for 700~mK, we obtain
$U_{0} \sim 16$~K.
At this field and temperature, $\nu \sim 4$,
leading to $\varepsilon \sim 0.005$.

Equation~\ref{comp} also shows that the value of $\omega\tau$
contributes to the complex resistivity.
Taking a frequency $\omega/2 \pi \approx 25$~MHz from the current
experiments and
$\tau$ from measurements of the normal-state
resistivity~\cite{calculation} (or the penetration
depth: see Figure~\ref{lamvt}), we obtain $\omega\tau \sim 0.005$.
Therefore, the fact that both $\varepsilon$ and $\omega \tau$
are $\ll 1$ indicates that the
surface-impedance and skin-depth effects
have negligible impact~\cite{cnc} on the experiments
on \lbets ~reported in this work.



\section{References}
\begin{thebibliography}{9}
\bibitem{review}
J. Singleton, Reports on Progress in Physics {\bf 63}, 1111 (2000).
\bibitem{carrington}
A. Carrington, I.J. Bonalde, R. Prozorov, R.W. Gianetta, A.M. Kini,
J. Schlueter, H.H. Wang, U. Geiser and J.M. Williams,
Phys. Rev. Lett., {\bf 83},
4172 (1999).
\bibitem{elsinger}
H. Elsinger, J. Wosnitza, S. Wanka, J. Hagel,
D. Schweitzer and W. Strunz, Phys. Rev. Lett. {\bf 84}, 6098 (2000).
\bibitem{tunnelling}
K.~Ichimura {\it et al.}, {\it Synth. Met.}
{ \bf 103}, 1812 (1999);
{\it Journal-of-Superconductivity}, {\bf 12}, 519 (1999);
T. Arai {\it et al.}, Phys. Rev. B {\bf 63}, 104518 (2001).
\bibitem{french}
S. Lefebvre, P. Wzietek, S. Brown, C. Bourbonnais, D. Jerome, C. Meziere,
M. Fourmigue and P. Batail, Phys. Rev. Lett. {\bf 85}, 5420 (2000).
\bibitem{schmalian}
J\"{o}rg Schmalian, Phys. Rev. Lett.
{\bf 81}, 4232 (1998).
\bibitem{aoki}
Kazuhiko Kuroki and Hideo Aoki,
Phys. Rev. B {\bf 60}, 3060 (1999).
\bibitem{maki}
K.~Maki, E.~Puchkaryov, H.~Won,
{\it Synth.~Met.} {\bf 103}, 1933 (1999).
\bibitem{charffi}
R.~Louati, S.~Charfi-Kaddour, A.~Ben~Ali, R.~Bennaceau
and M.~Heritier,
Synthetic Metals {\bf 103}, 1857 (1999).
\bibitem{betsreview1}
R.~Kato, H.~Kobayashi and A. Kobayashi,
Synth. Met. {\bf 42}, 2093 (1991).
\bibitem{growthdetails}
L. K. Montgomery, T. Burgin, J. C. Huffman, K. D. Carlson, J. D.
Dudek, G. A. Yanconi, L. A. Menga, P. R. Mobley, W. K. Kwok, J. M.
Williams, J. E. Schirber, D. L. Overmyer, J. Ren, C. Rovira, and M.
-H. Wangbo, Synth. Met. \textbf{56} 2090 (1993).
\bibitem{calculation}
L. K. Montgomery, T. Burgin and J. C. Huffman, J. Ren and M. -H.
Whangbo, Physica C \textbf{219}, 490 (1994).
\bibitem{grow2}
H. Kobayashi, T.~Udagawa, H. Tomita, K. Bun, T. Naito
and A. Kobayashi, Chem. Lett. {\bf 1993}, 1559 (1993)
\bibitem{ujiuji}
Uji {\it et al.} have recently observed what is believed
to be magnetic field-induced superconductivity in
$\lambda$-(BETS)$_2$FeCl$_4$ (see S.~Uji {\it et al.}, Nature,
{\bf 410}, 908 (2001)). However, this salt is not a superconductor
at zero field.
\bibitem{whangbo}
M.-H.~Whangbo, H.-J.~Koo and L.K.~ Montgomery, to be published (2001).
\bibitem{montya}
L.K. Montgomery, , T. Burgin, T. Miebach, D. Dunham and J.C.
Huffman, Mol. Cryst. Liq. Cryst. {\bf 284}, 73(1996).
\bibitem{pulsed}
F. Herlach, Reports on Progress in Physics,
{\bf 62}, 859 (1999).
\bibitem{house}
A. A. House, N. Harrison, S. J. Blundell, I. Deckers,
J. Singleton, F. Herlach, W. Hayes, J. A. A. J. Perenboom,
M. Kurmoo and P. Day,
{\it Phys. Rev. B} {\bf 53}, 9127 (1996).
\bibitem{msamro}
M.S. Nam, S.J. Blundell. A. Ardavan, J.A. Symington
and J. Singleton, J. Phys: Condens. Matter {\bf 13}, 2271 (2001)
\bibitem{shoenberg}
{\it Magnetic oscillations in metals}
D. Shoenberg, (Cambridge
University Press, 1984).
\bibitem{caulfield}
J.M.~Caulfield, W.~Lubczynski, F.L.~Pratt,
J.~Singleton, D.Y.K.~Ko,
W.~Hayes, M.~Kurmoo and P.~Day,
J. Phys.: Condens. Matter {\bf 6}, 2911 (1994).
\bibitem{QBB}
K.F. Quader, K.S. Bedell, G.E. Brown, Phys. Rev. B
{\bf 36}, 156 (1987); A.J. Leggett, Annals of Physics {\bf 46}, 76
(1968); W. Kohn, Phys. Rev. {\bf 123}, 1242 (1961);
K.~Kanki and K.~Yamada, J.~Phys. Soc. Jpn.
{\bf 66}, 1103 (1997).
\bibitem{tossnitza}
{\it Fermi surfaces of low-dimensional
organic metals and superconductors},
J.~Wosnitza (Springer-Verlag, Berlin, 1996)
\bibitem{kappabets}
C. H. Mielke, N. Harrison, D. G. Rickel, A. H. Lacerda,
R. M. Vestal, and L. K. Montgomery, Phys. Rev. B \textbf{56}, R4309
(1997).
\bibitem{eva}
N.~Harrison, E.~Rzepniewski,
J.~Singleton, P.J.~Gee, M.M. Honold,
P.~Day and M.~Kurmoo,
J. Phys.: Condens. Matter, {\bf 11}, 7227 (1999).
\bibitem{montxray}
Unpublished results of L.K. Montgomery and J.C. Huffman (2001).
\bibitem{mck}
For a discussion of the issue of interplane dispersion
in quasi-two-dimensional metals and
measurements such as this one, see
P.~Moses and R.H.~McKenzie, Phys. Rev. B {\bf 60}, 7998 (1999); D.~Yoshioka,
J.Phys. Soc. Jpn {\bf 64}, 3168 (1995);
R.H.~McKenzie and P.~Moses,
Phys. Rev. Lett.{\bf 81}, 4492 (1998); Phys. Rev. B {\bf 60}, 11241 (1999).
\bibitem{hanasaki}
T.~Osada {\it et al.}, Phys. Rev. Lett. {\bf 77}, 5261 (1996);
N. Hanasaki {\it et al.}, Phys. Rev. B {\bf 57}, 1336 (1998);
{\it ibid.} {\bf 60}, 11210 (1999)
\bibitem{russian}
Others propose that ``self-crossing orbits'' are more
effective than closed orbits in averaging $v_{\perp}$;
however, the geometrical constraints on the peak
in $\rho_{zz}$ are identical. See
V.G.~Peschansky and M.V. Kartsovnik, Phys. Rev. B {\bf 60},
11207 (1999); I.J.~Lee and M.J.~Naughton, Phys. Rev. B {\bf 57}, 7423 (1998).
\bibitem{goddard}
J. Singleton, P.A. Goddard, A. Ardavan, N. Harrison, S.J. Blundell,
J.A. Schlueter and A.M. Kini, cond-mat/0104570 (2001),
Phys. Rev. Lett., submitted;
P. Goddard, A. Ardavan, J. Singleton and J.A.~Schlueter,
to be published.
\bibitem{ashcroft}
{\it Solid State Physics},
by N.W. Ashcroft and N.D. Mermin
(Holt-Saunders, New York, 1976).
\bibitem{footies}
Alternative methods, such as the fitting of
more complex functions to the background magnetoresistance
and peak yielded negligible gains in accuracy
and reproducibility, and were more time-consuming.
\bibitem{footnote22}
The value $t_{\perp}=0.21$~meV assumes that the interplane
distance $b$ which dominates the interlayer
transport is that given by X-ray crystallography, 18.4~\AA.
\bibitem{tdo}
C.T. VanDeGrift, Rev. Sci. Inst., {\bf 46}, 599 (1975);
C.T. VanDeGrift, and D.P. Love, Rev. Sci. Inst. {\bf 52}, 712 (1981).
\bibitem{janeloff}
J. Singleton, J.A. Symington, M.S. Nam, A. Ardavan,
M.~Kurmoo and P. Day,
J. Phys.: Condens Matter
{\bf 12} L641 (2000).
\bibitem{bleaney}
See {\it e.g.} {\it Electricity and Magnetism},
B.I.~Bleaney and B.~Bleaney, Oxford University Press
(third edition, Oxford, 1990).
\bibitem{mielkethesis}
C.H. Mielke, PhD Thesis, Clark University 1996.
\bibitem{footnote1}It should be emphasised
that this measurement technique represents a very weak perturbation
of the sample.
The radio-frequency (rf) magnetic
field in the coil is $\sim 1~\mu$T, several orders of magnitude smaller
than typical applied fields.
Moreover, the data
were taken at a frequency of 25 MHz, whilst a BCS
estimate of the pair-breaking frequency
in \lbets~ is $\sim 160$~GHz~\cite{mielkethesis}, {\it i.e.}
the measurement frequency is insufficient to cause
pair-breaking.
\bibitem{schrama}
J.M.~Schrama, J. Singleton, R.S. Edwards,
A. Ardavan, E. Rzepniewski, R. Harris, P. Goy,
M. Gross, J. Schlueter, M. Kurmoo and P. Day,
J. Phys.: Condens. {\bf 13} 2235 (2001)
\bibitem{hill}
S.~Hill, Phys. Rev. B {\bf 62},8699 (2000).
\bibitem{wu}
D. Wu and S. Sridhar, Phys. Rev. Lett.,  {\bf 65} 2074 (1990).
\bibitem{abrikosov}
A. A. Abrikosov, Zh. Eksperim. i Teor. Fiz.  {\bf 32} 1442 (1957).
\bibitem{tinkham}
M.~Tinkham, {\it Introduction to superconductivity, 2nd ed.},
(McGraw Hill, 1996).
\bibitem{bs}
J. Bardeen and M. J. Stephen, Phys. Rev. \textbf{140}
A1197 (1965).
\bibitem{gorter}
C. J. Gorter and H. B. G. Casimir, Phys. Z.  {\bf 35} 963 (1934).
\bibitem{campbell}
A.M. Campbell and J.E. Evetts,
Adv. Phys. {\bf 21}, 199 (1972).
\bibitem{cnc}
M. W. Coffey and J. R. Clem, Phys. Rev. Lett.,  {\bf 67} 386 (1991).
\bibitem{hcpapers}
H.~Ito, T.~Ishiguro, T.~Komatsu, G.~Saito and H.~Anzai,
Physica B {\bf 201}, 470 (1994) and references therein;
T.~Ishiguro, H.~Ito, Yu.V.~Sushko, A.~Otsuka and G.~Saito,
Physica B {\bf 197}, 563 (1994);
F.~Zuo, X.~Su, P.~Zhang,
J.A.~Schleuter, M.E.~Kelly and J.M.~Williams,
Phys. Rev. B {\bf 57} R5610 (1998);
F.~Zuo, J.A.~Schleuter, M.E.~Kelly
and J.M.~Williams, Phys. Rev. B {\bf 54},
11973 (1996).
\bibitem{belin}
S.~Belin, T.~Shibauchi, K.~Behnia and T.~Tamegai,
J. of Superconductivity, {\bf 12}, 497 (1999).
\bibitem{finnemore}
D. K. Finnemore, {\it Phenomenology  and Applications  of High Temperature
Superconductors},
Addison Wesley, Reading MA (1992), pp. 164.
\bibitem{boeb}
Y. Ando, G.S. Boebinger, A. Passner, L.F. Schneemeyer,
T. Kimura, M. Okuya, S. Watauchi, J. Shimoyama, K. Kishio,
K. Tamasaku, N. Ichikawa and S. Uchida, Phys. Rev. B
{\bf 60}, 12475 (1999) and references therein.
\bibitem{houghton}
A. Houghton, R. A. Pelcovits, and A. Sudbo, Phys.
Rev. B, 40 (1989) 463.
\bibitem{saturationbombing}
See {\it e.g.} {\it Principles of the theory of solids},
(second edition) J.M.~Ziman (Cambridge University Press, 1972)
and references therein.
\bibitem{mielkenew}
C.H. Mielke {\it et al.} to be published.
\bibitem{jscm}
J. Singleton and C.H. Mielke, to be published (2001).
\bibitem{sasaki}
T.~Sasaki, W.~Biberacher, K. Neumaier, W. Hehn, K. Andres and T.~Fukase,
Phys. Rev. B. {\bf 57}, 10889 (1998).
\bibitem{lang}
M. Lang, F. Steglich, N. Toyota and T. Sasaki,
Phys. Rev. B {\bf 49}, 15227 (1994).
\bibitem{leepratt}
S.L. Lee, F.L. Pratt, S.J. Blundell, C.M. Aegerter, P.A. Pattenden,
K.H. Chow, E.M. Forgan, T. Sasaki, W. Hayes and H. Keller,
Phys. Rev. Lett. {\bf 79}, 1563 (1997).
\bibitem{mola}
M. Mola, S. Hill, J.S. Brooks and J.S. Qualls, Phys. Rev. Lett.
{\bf 86}, 2130 (2001).
\bibitem{nmrnmr}
H. Mayaffre, P. Wzietek, D. Jerome and S. Bazovskii,
Phys. Rev. Lett. {\bf 76}, 4951 (1996).
\bibitem{graebner}
See J.E. Graebner, R.C. Haddon, S.V. Chichester and S.H. Glarum,
Phys. Rev. B {\bf 41}, 4808 (1990) and references therein.
\bibitem{graybeal}
A similar curvature in $H_{\rm c2}$ versus $T$ data has been
observed in artificially-constructed 2D superconductors
(Mo-Ge films of thickness $\sim 2$~nm; the field was applied
perpendicular to the 2D planes), suggesting that
it is a generic property of 2D superconductors; see
J.M. Graybeal and M.R. Beasley, Phys. Rev. B {\bf 29},
4167 (1984).
\bibitem{ld}
W. E. Lawrence and S. Doniach, in E. Kanda (ed.)  Proc.
 12th Int.  Conf.  Low Temp.  Phys.  (Kyoto, 1970), pp. 361.
\bibitem{naka}
T.~Nakamura, T.~Komatsu,
G.~Saito, T.~Osada, S.~Kagoshima, N.~Miura,
K.~Kato, Y.~Maruyama and K.~Oshima,
J.~Phys. Soc. Jpn., {\bf 62} 4373 (1993).
\bibitem{Morris}
R.~C,~Morris, R.~V.~Coleman, and R.~Bhandari,
Phys. Rev. B  {\bf 5}, 895 (1972).
\bibitem{guelph}
This procedure was carried out using Equations~\ref{xixixi}, \ref{gl},
and \ref{xixixixi} in the limit
$T \rightarrow 0$~\cite{mielkenew}; see
Section~\ref{discussion} for a discussion of some
of the difficulties inherent in this procedure.
\bibitem{parks}
{\it Superconductivity}, edited by R.D.~Parks
(Marcel Dekker, New York, 1969).
\bibitem{notethis}
A number of authors have compared the magnetic-field orientation
dependence of $H_{\rm c2}$ in organic superconductors
with various theoretical expressions. It is found that
the Ginzberg-Landau-effective-mass expressions fit the data reasonably
well for angles away from $\theta=90^{\circ}$. However,
close to in-plane magnetic fields ($\theta=90^{\circ}$),
mechanisms limited by
spin (rather than orbital) effects take over, so that a
simple ratio of the in-plane $H_{\rm c2}$ to the perpendicular
$H_{\rm c2}$ is NOT a reliable guide to the relative
sizes of the coherence lengths.
This point is discussed at some length in Reference~\cite{review},
in M.-S. Nam {\it et al.}
J. Phys.: Condens. Matter, {\bf 11}, L477 (1999), Reference~\cite{janeloff}
and References cited in these papers.
In Reference~\cite{mielkenew} these problematic orientations are
deliberately avoided.
\bibitem{white}
W.R. White, A.~Kapitulnik and M.R. Beasley,
Phys. Rev. Lett. {\bf 66}, 2826 (1991).
\bibitem{perpnote}
The majority of dimensional crossover studies
employing artificial layered structures focus
on the effect of an in-plane magnetic field,
which is not relevant in the current context;
see {\it e.g.} S.T Ruggiero,
T.W. Barbee and M.R. Beasley, Phys.Rev. Lett.
{\bf 45}, 1299 (1980) and references therein.
\end{thebibliography}
\end{document}